\documentclass[twocolumn,showpacs,preprintnumbers,amsmath,amssymb]{revtex4}

\usepackage{graphicx}
\usepackage{dcolumn}
\usepackage{bm}
\usepackage{color}

\begin{document}


\title{First-order phase transition and anomalous hysteresis of Bose gases in optical lattices}

\author{Daisuke Yamamoto$^{1}$}
\email{d-yamamoto@riken.jp}
\author{Takeshi Ozaki$^2$}
\author{Carlos A. R. S\'a de Melo$^{3}$}
\author{Ippei Danshita$^{4,5}$}
\affiliation{
{$^1$Condensed Matter Theory Laboratory, RIKEN, Wako, Saitama 351-0198, Japan}
\\
{$^2$Faculty of Business Administration and Information,
Tokyo University of Science, Suwa, Chino, Nagano 391-0292, Japan}
\\
{$^3$School of Physics, Georgia Institute of Technology, Atlanta, Georgia 30332, USA}
\\
{$^4$Yukawa Institute for Theoretical Physics, Kyoto University, Kyoto 606-8502, Japan}
\\
{$^5$Computational Condensed Matter Physics Laboratory, RIKEN, Wako, Saitama 351-0198, Japan}
}
\date{\today}

\begin{abstract}
We study the first-order quantum phase transitions of Bose gases in optical lattices. A special emphasis is placed on an anomalous hysteresis behavior, in which the phase transition occurs in a unidirectional way and a hysteresis loop does not form. We first revisit the hardcore Bose-Hubbard model with dipole-dipole interactions on a triangular lattice to analyze accurately the ground-state phase diagram and the hysteresis using the cluster mean-field theory combined with cluster-size scaling. Details of the anomalous hysteresis are presented. We next consider the two-component and spin-1 Bose-Hubbard models on a hypercubic lattice and show that the anomalous hysteresis can emerge in these systems as well. In particular, for the former model, we discuss the experimental feasibility of the first-order transitions and the associated hysteresis. We also explain an underlying mechanism of the anomalous hysteresis by means of the Ginzburg-Landau theory. From the given cases, we conclude that the anomalous hysteresis is a ubiquitous phenomenon of systems with a phase region of lobe shape that is surrounded by the first-order boundary.
\end{abstract}

\maketitle
\section{\label{1}Introduction} 
Systems of ultracold gases confined in optical lattices have been often used as quantum simulators for the studies of strongly interacting many-body physics~\cite{bloch-12}. It is of great advantage that many essential features, including the statistics of particles, the optical-lattice depth, the filling factors, and the interparticle interactions, are widely controllable. The ultimate goal of the quantum simulation is to emulate the complex models that cannot be accurately solved with currently available theoretical approaches, such as the Fermi-Hubbard model and the quantum spin models on a kagome lattice. However, to that end, it is also important to examine in advance the performance of the optical-lattice systems as simulators by comparing experiments with theories in solvable models. Trotzky {\it et al.}~has indeed presented quantitative comparison between the experiment with spinless Bose gases in a cubic optical lattice and the quantum Monte Carlo (QMC) analyses on the corresponding Bose-Hubbard model to show that they agree well with each other regarding phase transitions from superfluid (SF) to normal at finite temperatures upon approaching the Mott transition point~\cite{trotzky-10}.

The recent development of experimental techniques has offered possibilities to simulate more complicated and richer systems. For example, mixing different hyperfine states~\cite{anderlini-07, trotzky-08, widera-08, weld-09, gadway-10, chin-06, jordens-08, schneider-08, taie-12}, atomic species~\cite{catani-08, gunter-06, ospelkaus-06}, or isotopes~\cite{sugawa-11} in optical lattices has added pieces to access a variety of quantum phases and phase transitions. Moreover, a long-range and anisotropic interaction between particles has been introduced by the creation of degenerate gases of atoms with strong magnetic dipole-dipole interactions, such as chromium~\cite{griesmaier-05}, dysprosium~\cite{lu-11}, and erbium~\cite{aikawa-12}. Experimentalists have also attempted to realize the quantum degeneracy in gases of heteronuclear polar molecules with stronger dipolar interactions, such as KRb~\cite{ni-08,aikawa-10} and LiCs~\cite{deiglmayr-08}. It is also worth noting that even the lattice geometry is flexibly controllable; two-legged ladder~\cite{anderlini-07, trotzky-08, sebby-06, folling-07, yachen-10}, triangular~\cite{becker-10,struck-11}, honeycomb~\cite{soltan-11,soltan-12,tarruell-12}, and kagome~\cite{jo-12} optical lattices have been created in recent experiments in addition to the standard hypercubic lattices~\cite{greiner-02, stoeferle-04, spielman-07}. Such rapid expansion of the range of application demands further sophistication of the performance of quantum simulators.

More specifically, it has been predicted that there exist many quantum phase transitions of first order in some of the systems described above~\cite{altman-03, kuklov-04, chen-10, ozaki-12, krutitsky-04, krutitsky-05, kimura-05, kimura-06, pai-08,deforges-11, murthy-97, wessel-05, boninsegni-05, heidarian-05, melko-05, sen-08, heidarian-10, pollet-10, yamamoto-12a, bonnes-11, zhang-11, danshita-09, batrouni-00, yamamoto-12b}. Hence, the quantum simulator has to be able to distinguish a first-order (discontinuous) phase transition from a second-order (continuous) one in order to map out correctly the phase diagrams of those systems. Indeed, the first-order transition between two phases with different pseudo-spin orders has been experimentally identified by measuring a discontinuous jump of the momentum distribution in the system of Bose gases confined in a triangular optical lattice~\cite{struck-11}. Furthermore, since the first-order transition is one of the fundamental subjects in thermodynamics and statistical physics, the observability of the accompanying phenomena such as hysteresis can be a touchstone to test the performance of a quantum simulator based on trapped atomic/molecular gases in optical lattices.

The first-order quantum phase transitions in the Bose-Hubbard systems are interesting in part because they occasionally exhibit an anomalous hysteresis behavior as predicted for the melting transition of the hardcore Bose-Hubbard model with long-range interactions on a triangular lattice~\cite{yamamoto-12a}. In this anomalous hysteresis, the phase transition occurs only unidirectionally; while the solid order melts when the chemical potential varies, the system in the SF state cannot be solidified again in the inverse process. However, comprehensive understanding of this phenomenon has not been achieved yet in the sense that the following two questions are open. First, what are the necessary conditions for the anomalous hysteresis to emerge? This question can be inductively addressed if one shows a few other examples. Second, can the hysteresis be described within the Ginzburg-Landau theory for the first-order transitions? Answering the second question is important because the anomalous hysteresis seemingly contradicts with the Ginzburg-Landau theory that always predicts the spinodal points upon varying monotonically the coefficient relevant to the transition.

In this paper, we study quantum phases and phase transitions of Bose gases in optical lattices, with a central focus on the first-order transitions.
We consider the systems of (i) dipolar hardcore bosons on a triangular lattice and (ii) multi-component bosons on a hypercubic lattice. Both systems are known to exhibit a first-order phase transition. We discuss the first-order transition phenomena of these systems, including the possibility of experimental observations and the expected characteristic hysteresis behavior, in a comprehensive manner.

In the previous work on system (i), Yamamoto {\it et al.}~mainly analyzed a simple model in which the infinite-range dipole-dipole interaction is truncated to be only nearest-neighbor (NN) interaction~\cite{yamamoto-12a}. We first reconsider in more detail the first-order transitions for the model with full dipole-dipole interaction using a larger-size cluster mean-field (CMF) method (with up to 15-site clusters) and the cluster-size scaling. We quantitatively evaluate the phase boundaries through the size-scaling of the CMF data and predict the presence of several additional phases that have not been reported in the previous QMC simulations~\cite{pollet-10}. Besides, we find that the untruncated model can also exhibit the anomalous hysteresis. System (i) has several special features, namely, two-dimensionality, geometric (classical) frustration, tripartite lattice, hardcore nature, long-range interactions, and supersolidity. In order to identify which features are essential for the anomalous hysteresis, we next examine much simpler models that do not have those special features, i.e., system (ii): multi-component softcore bosons on a hypercubic lattice.

As for system (ii), we specifically investigate the first-order phase transition from SF to the Mott insulator (MI) in the two-component and spin-1 Bose-Hubbard models within the site-decoupling mean-field approximation. We show that this first-order transition exhibits the anomalous hysteresis behavior as well as in system (i). The two-component model has been already emulated in several previous experiments~\cite{anderlini-07, widera-08, trotzky-08, weld-09, gadway-10, catani-08}. Having these experiments in mind, we discuss what conditions are required for observing the hysteresis associated with the first-order transition. Especially, we calculate the transition temperature from SF to normal phase upon approaching the Mott transition, in order to reveal the requirements regarding the temperature and the lattice depth. It will be shown that the required temperature is within the reach of current (or near-future) experimental techniques and, at least, much higher than the temperature necessary to observe magnetism in optical-lattice systems. Moreover, taking advantage of the simplicity of the two-component model with respect to the first-order transition, we construct the Ginzburg-Landau theory for the anomalous hysteresis to elucidate its physical mechanisms. Extracting common features from the obtained examples, we finally argue that the anomalous hysteresis emerges universally (independently from the frustration and the dimensionality) when a phase region of lobe shape is surrounded by a first-order boundary.

The remainder of the paper is organized as follows. In Sec.~\ref{2}, we investigate the hardcore Bose-Hubbard model with long-range dipole-dipole interaction on a two-dimensional triangular lattice. The ground-state phase diagram obtained from the cluster-size scaling of the CMF data is compared with the previous QMC data in Sec.~\ref{2-1} and the hysteresis behavior accompanying first-order transitions is discussed in Sec.~\ref{2-2}. In Sec.~\ref{3}, we analyze the two-component Bose-Hubbard model on a hypercubic lattice. We discuss the first-order SF-to-MI transitions induced by varying the hopping amplitude or the chemical potential. After the site-decoupling mean-field theory is introduced in Sec.~\ref{3-1}, the cases of equal and unequal intra-component interactions are studied in Sec.~\ref{3-2} and Sec.~\ref{3-3}. 
In Sec.~\ref{3-4}, we construct the Ginzburg-Landau theory for the anomalous hysteresis phenomenon. 
In Sec.~\ref{4}, the spin-1 Bose-Hubbard model on a hypercubic lattice is analyzed.
The results are summarized in Sec.~\ref{5}.

\section{\label{2} Dipolar bosons in a triangular lattice}
We consider a dipolar Bose gas loaded into a deep triangular optical lattice and describe the system with the following Bose-Hubbard model,
\begin{eqnarray}
\hat{H}=
-J\sum_{\langle j,l \rangle}
(\hat{a}^{\dagger}_{j} \hat{a}_{l}+{\rm H.c.})
+\frac{1}{2}\sum_{ j, l }
V_{jl} \hat{n}_{j}\hat{n}_{l}
-\mu \sum_j \hat{n}_j,
\label{hamiltonian}
\end{eqnarray}
where $\hat{a}^{\dagger}_{j}$ and $\hat{n}_j=\hat{a}^{\dagger}_{j} \hat{a}_{j}$ are the creation and number operators of the hardcore bosons at site $j$, $J$ denotes the hopping amplitude between NN pairs, and $\mu$ the chemical potential. Here we take the hardcore boson limit, which means that two or more bosons are not allowed to occupy the same site due to the strong on-site interaction. 
We assume that the dipole moments are polarized by the external field in the direction perpendicular to the lattice plane such that the dipole-dipole interaction $V_{jl}$ is well approximated as~\cite{goral-02}
\begin{eqnarray}
V_{jl}&=&\left\{ \begin{array}{ll}
Vd^3/\left|{\bf r}_{j}-{\bf r}_{l}\right|^3   &  (j\neq l) \\
0 & (j=l)
\end{array} \right.,\label{Dip}
\end{eqnarray}
where $d$ is the lattice spacing and ${\bf r}_j=(j_x d,j_y d)$ with integers $j_x$ and $j_y$ is a lattice vector at site $j$.

Since the dipole-dipole interaction rapidly falls off as the inverse cube of the distance $\left|{\bf r}_{j}-{\bf r}_{l}\right|$, several essential properties may be captured using a simplified model with keeping only the NN repulsion $V$. Such a truncated model has been the subject of intensive study~\cite{murthy-97,wessel-05,boninsegni-05,heidarian-05,melko-05,sen-08,heidarian-10,yamamoto-12a,bonnes-11,zhang-11} due to its simplicity. The outline of the ground-state phase diagram for this model has been given by the earlier MF study~\cite{murthy-97} and QMC calculations~\cite{wessel-05}. Recently, analyses with a large-size CMF approach have pointed out that the transition between the SF and supersolid (SS) phases should be of first order~\cite{yamamoto-12a}, and this has been reconfirmed by the latest QMC analyses~\cite{bonnes-11,zhang-11}. Thus, the complete ground-state phase diagram is well established for the truncated model.

For the case of full dipole-dipole interaction given in Eq.~(\ref{Dip}), only a few previous works~\cite{pollet-10,yamamoto-12a} have calculated the ground-state phase diagram. The QMC analyses in Ref.~\onlinecite{pollet-10} have shown that there exist the two solid phases with one-third and two-third fillings, and the SS phase in between the two solids, as in the case of the truncated model. Neither additional solids nor SSs were found.
However, the error bars attached to the phase boundaries are rather large in the QMC data of Ref.~\onlinecite{pollet-10} so that some of important properties of the phase diagrams might have been missed. Hence, more careful analyses are needed for further refinement of the phase diagram.

In general, simulations on long-range interacting systems are much more difficult compared to the case of short-range models. The hardcore-boson model can be treated as a system of $S=1/2$ spins. In the simulation on the $N$-spin system with long-range interactions, one has to consider $\mathcal{O}(N^2)$ different pairs of spins, which makes the computation time per Monte Carlo sweep grow quite rapidly as the system size increases. One way to overcome the problem is to employ the latest QMC algorithm, called the $\mathcal{O}(N)$ cluster Monte-Carlo method~\cite{fukui-09}, which treats efficiently long-range interactions. In the following of this section, we attack the problem using another numerical approach based on the exact-diagonalization techniques with a self-consistent mean-field ansatz.

\subsection{\label{2-1} Ground-state phase diagram}
We re-examine the ground-state phase diagram of the dipolar Bose-Hubbard model by means of the large-size CMF method and the cluster-size scaling analysis (CMF+S)~\cite{yamamoto-12b}. We use the series of the clusters that consist of $N_{\rm C}=$ 3, 6, 10, and 15 sites. The shape of each cluster is depicted in Table~\ref{table1}. 
In the CMF method, we perform exact diagonalizations of the cluster system with mean-field boundary condition. The values of the mean fields are determined by solving the CMF self-consistent equations (see Ref.~\onlinecite{yamamoto-12b} for more details). 
Furthermore, we carry out a cluster-size scaling of the CMF data with the scaling parameter $\lambda$ defined by $N_{\rm B}/(N_{\rm C}\times z/2)$, 
where $N_{\rm B}$ is the number of bonds within the cluster and $z = 6$ is the coordination number of the triangular lattice.

\begin{table}[b]
\caption{\label{table1}The series of the clusters used in the CMF calculations. The values of $N_{\rm B}$ and $\lambda$ are also listed. }
\includegraphics[scale=0.5]{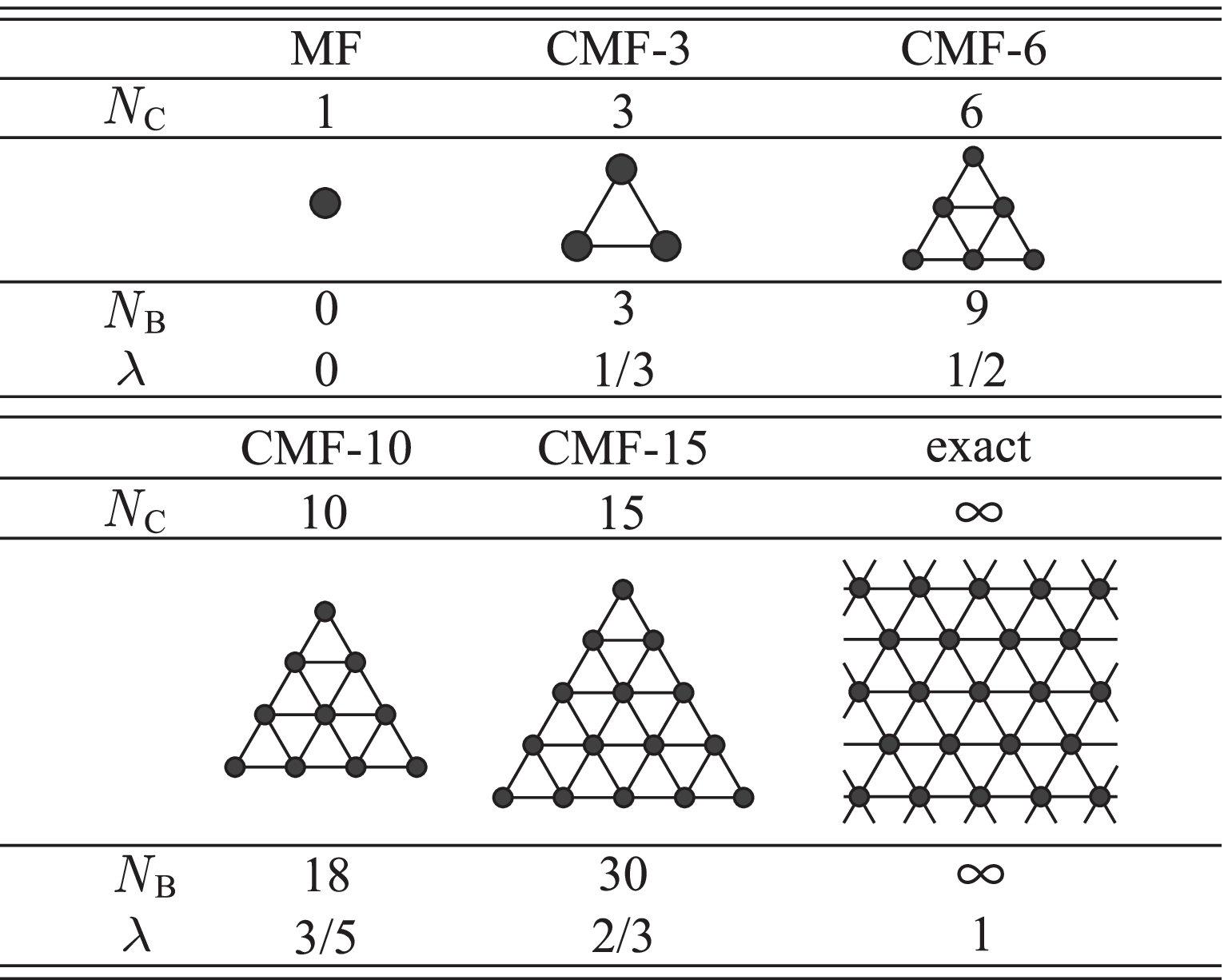}
\end{table}

\begin{figure}[t]
\includegraphics[scale=0.48]{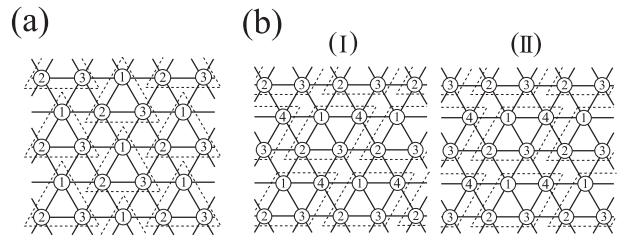}
\caption{\label{fig2}
(a) A three-sublattice $\sqrt{3}\times\sqrt{3}$ structure and (b) two types of four-sublattice structures assumed in the present calculations. The lattice sites with the same number belong to the same sublattice. 
}
\end{figure}

In the CMF calculation, one assumes background sublattice structures that are expected to emerge from the given lattice geometry and the nature of interactions~\cite{yamamoto-12b}. In the simple truncated model, only the states with $\sqrt{3}\times \sqrt{3}$ sublattice structure [Fig.~\ref{fig2}(a)] emerge as a natural result of the interplay between the NN density-density repulsion and the geometry of the triangular lattice~\cite{murthy-97,wessel-05,boninsegni-05,heidarian-05,melko-05,sen-08,heidarian-10,yamamoto-12a,bonnes-11,zhang-11}. In the case of full dipole-dipole interaction, more complicated patterns that have a larger unit cell are also expected to be formed since the interaction can act on longer-distant boson pairs~\cite{sansone-10a}. Therefore, we take into account two types of four-sublattice patterns shown in Fig.~\ref{fig2}(b) in addition to the basic three-sublattice $\sqrt{3}\times \sqrt{3}$ structure.

\begin{figure}[t]
\includegraphics[scale=0.33]{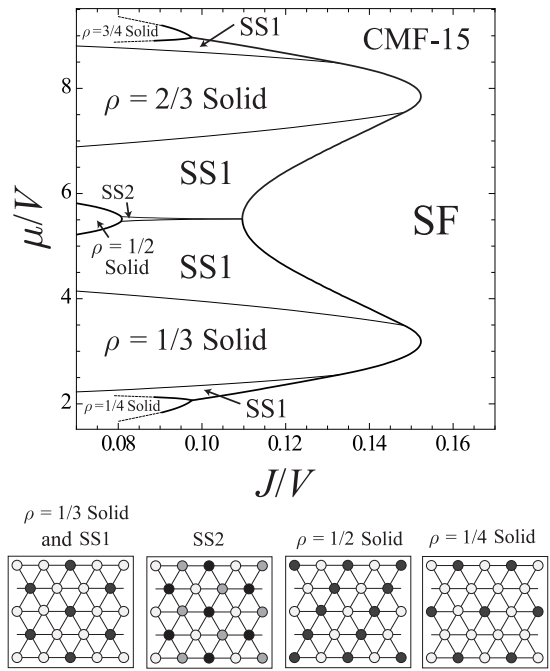}
\caption{\label{fig1}
Ground-state phase diagram of dipolar bosons in a triangular lattice obtained by the CMF-15. Second- and first-order phase transitions are indicated by the thin and thick lines, respectively. 
Lower panels: Schematic pictures of the density patterns in each phase. Because of the particle-hole symmetry, we show only the phases on the low-density side of the phase diagram. 
}
\end{figure}

Figure~\ref{fig1} shows the ground-state phase diagram of the dipolar Bose-Hubbard model given in Eq.~(\ref{hamiltonian}) obtained by the CMF method with the 15-site cluster (CMF-15). The phase diagram is symmetric around $\mu/V = (\mu/V)_0\equiv \sum_{l}V_{jl}/2V\approx 5.517$, reflecting the particle-hole symmetry of the hardcore boson system. 
We see the solids with the filling factor $\rho=1/4$, $1/3$, $1/2$, $2/3$, and $3/4$, and the two types of SS, which we name SS1 and SS2, as well as the standard SF state. 
While the solid phases form a superlattice pattern in the local density $\langle \hat{n}_j\rangle$, i.e., a crystalline order, the SF state is identified by the order parameter $\Psi\equiv \sum_j \langle \hat{a}_j\rangle /M$, where $M$ denotes the number of lattice sites. 
The SS phases simultaneously possess the SF order and the crystalline order. 
To determine the boundary lines of first-order transitions, we used the Maxwell construction in ($J/V,\chi$)-plane, where $\chi\equiv \sum_{\langle j,l \rangle}\langle\hat{a}^{\dagger}_{j} \hat{a}_{l}+\hat{a}^{\dagger}_{l} \hat{a}_{j}\rangle/M$. The filling factor is given by $\rho=\sum_j\langle \hat{n}_j\rangle/M$.

Solving the CMF self-consistent equations with the ansatz of sublattice structures shown in Figs.~\ref{fig2}(a) and~\ref{fig2}(b), we obtain the crystalline order of each solid or SS state depicted in the lower panels of Fig.~\ref{fig1}.
As shown in the phase diagram, the solids with $\rho=1/3$ and $2/3$, and SS1 are present at relatively large $J/V$, as in the truncated model~\cite{murthy-97,wessel-05,yamamoto-12a,bonnes-11,zhang-11}.
However, there also emerge the solids with $\rho=1/4$, $1/2$, and $3/4$, and SS2 due to the long-range nature of the dipole-dipole interaction. Moreover, many other solid and SS phases that cannot be described by the three-sublattice and four-sublattice ansatz should be found for smaller values of $J/V$, as in the case of a square lattice~\cite{sansone-10a}.

In the previous work~\cite{yamamoto-12a}, we presented the single-site mean-field (MF) and CMF-10 results of the ground-state phase diagram (see Fig.~4 in Ref.~\onlinecite{yamamoto-12a}). Compared to them, the solid and SS regions in Fig.~\ref{fig1} are shifted to the side of small $J/V$ and especially the region of SS2 phase is much narrower. This means that the results of the phase boundaries are not completely converged. In order to include systematically the effects of quantum fluctuations,
we carry out the cluster-size scaling of the CMF data with different sizes of clusters. In the CMF+S analysis, we perform linear fits of the data obtained from the three largest clusters, namely $N_{\rm C}=6$, $10$, and $15$. The advantage of the CMF+S analysis is that usually the series of CMF data are well fitted to the linear function even if the clusters are not quite large~\cite{yamamoto-12b}. This is attributed to the fact that the CMF method treats infinite-size systems by setting the mean-field boundary condition despite using a finite-size cluster. This point is an important difference from the standard exact diagonalization with open or periodic boundary condition. Indeed, the CMF+S procedure has successfully generated a quantitatively reliable result for the phase boundaries of the dipolar model on a square lattice~\cite{yamamoto-12b}. 

\begin{figure}[t]
\includegraphics[scale=0.33]{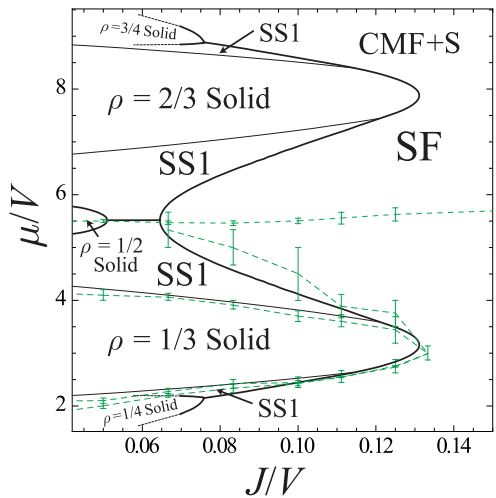}
\caption{\label{fig3}
(color online) The CMF+S result of the phase boundaries. For comparison, we also plot the QMC data extracted from Fig. 1 of Ref.~\onlinecite{pollet-10} (green dashed lines; see Ref.~\onlinecite{pollet-10} for details).  }
\end{figure}

Figure~\ref{fig3} shows the resulting scaled data of the phase boundaries in the ($J/V$,$\mu/V$)-plane.  
Although the solid and SS regions are shifted further to the side of small $J/V$, the relative location of the phases does not change even in the limit of $N_{\rm C}\rightarrow \infty$ except for the following point: The calculations with up to $N_{\rm C}=15$ clusters show that a narrow but finite region of the SS2 phase exists around half filling and the transition from the low-filling SS1 to high-filling SS1 state occurs continuously through the intermediate SS2 state (see Fig.~\ref{fig1}). However, the SS2 region rapidly shrinks with increasing the cluster size and completely disappears in the limit of $N_{\rm C}\rightarrow \infty$. Hence, the low-filling SS1 state is directly transformed into the high-filling SS1 state with a finite jump in, e.g., the filling factor at the particle-hole symmetric line $\mu/V = (\mu/V)_0$. This first-order nature of the SS1-SS1 transition has been predicted also for the truncated model~\cite{boninsegni-05,heidarian-05,sen-08,yamamoto-12a}.

\begin{figure}[t]
\includegraphics[scale=0.4]{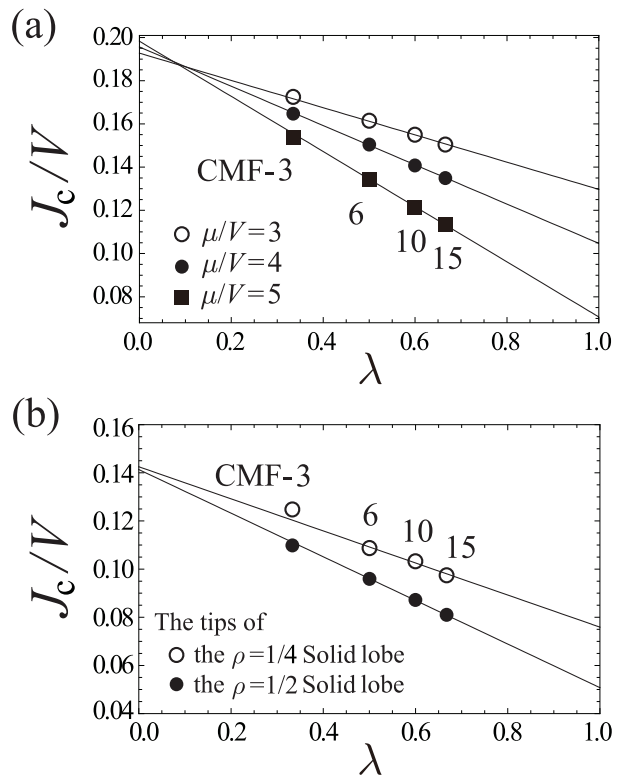}
\caption{\label{fig35}
(a) Examples of the cluster-size scaling of the CMF data for the phase boundaries $J_{\rm c}/V$ between the SF and SS1 phases (open symbols), and between the SF and $\rho=1/3$ solid phases (closed symbols). (b) Cluster-size scaling for the tips of the $\rho=1/4$ and $1/2$ solid lobes. }
\end{figure}

As shown in Fig.~\ref{fig35}(a), the linear fits of the CMF data for the SF-SS1 and $\rho=1/3$ solid-SS1 boundaries are fairly good. In fact, we can see in Fig.~\ref{fig3} that the scaled values and the QMC data of Ref.~\onlinecite{pollet-10} are compatible within the error bars. 
However, in contrast to Ref.~\onlinecite{pollet-10}, the SS1 phases are found not only in between the $\rho=1/3$ and 2/3 solids but also for $n<1/3$ and $n>2/3$. 
In addition, our CMF+S result predicts the existence of solid states with $\rho=1/4$, $1/2$, and $3/4$ in the plot range of the figure, although these states have not been detected in the QMC analysis for $J/V>0.045$~\cite{pollet-10}. As for the $\rho=1/2$ solid, the scaled phase boundary is reliable because the data of the three largest clusters almost lie in a straight line as seen in Fig.~\ref{fig35}(b). However, the linear fitting may not make a very good estimation for the $\rho=1/4$ solid boundary.
Therefore, more careful and precise analyses are required to corroborate the presence of the new phases. Such analyses may be able to be done with the use of the latest QMC algorithm~\cite{fukui-09,ohgoe-11} that can treat efficiently long-range interactions.

\subsection{\label{2-2} First-order phase transitions and anomalous hysteresis}
\begin{figure}[b]
\includegraphics[scale=0.37]{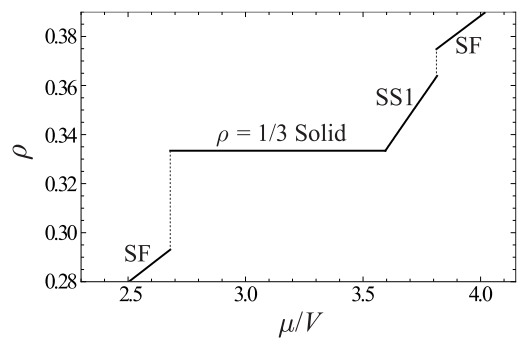}
\caption{\label{fig4}
The filling factor $\rho$ as a function of $\mu/V$ for $J/V=0.14$ (obtained by CMF-15). The first-order transition points are located at the discontinuous jumps in $\rho$. }
\end{figure}

The system of dipolar bosons in a triangular lattice exhibits first-order phase transitions indicated by the thick lines in Figs.~\ref{fig1}. The square-lattice model with dipole-dipole interactions also exhibits a first-order phase transition between the checkerboard solid and SF. However, the region of the first-order phase transition is limited only to the vicinity of the tip of the solid lobe~\cite{yamamoto-12b} and quite small. In contrast, one can find first-order phase transitions between SS and SF as well as between solid and SF in a broad region of the phase diagram of the triangular-lattice system. 
In Fig.~\ref{fig4}, we show the filling factor $\rho$ as a function of $\mu/V$ for $J/V=0.14$, which exhibits finite jumps at the first-order transition points. We focus on the quantum melting transition from solid to SF around the plateau with $\rho=1/3$. In Ref.~\onlinecite{yamamoto-12a}, it has been shown for the truncated model that the melting transition shows an anomalous hysteresis behavior. We establish here that the anomalous hysteresis can occur when the full dipole-dipole interaction is taken into account. 

\begin{figure}[t]
\includegraphics[scale=0.35]{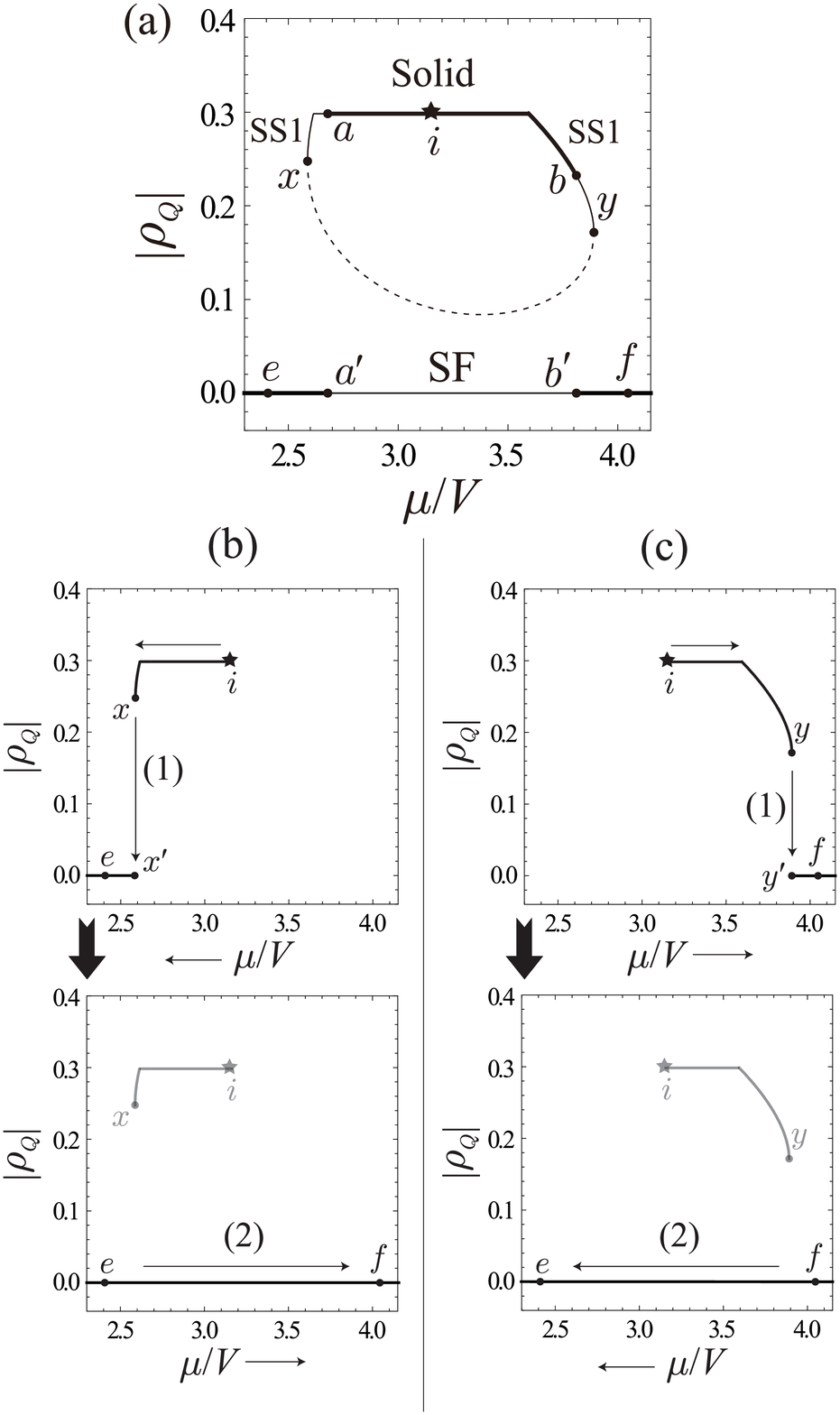}
\caption{\label{fig5}
(a) The solution curves of CMF-15 for the solid order parameter $|\rho_{\bf Q}|$ as a function of $\mu/V$ at zero temperature and $J/V=0.14$. The thick solid, thin solid, and dashed lines represent the ground, metastable, and unstable states, respectively. (b) and (c) The transition processes in the anomalous hysteresis. }
\end{figure}
In Fig.~\ref{fig5}(a), we plot the solid order parameter of the $\sqrt{3}\times\sqrt{3}$ sublattice structure, 
\begin{equation}
\rho_{\bf Q}= \frac{1}{M}\sum_j \langle \hat{n}_j\rangle e^{i{\bf Q}\cdot {\bf r}_j} ~{\rm with}~{\bf Q}=\left(\frac{4\pi}{3d},0\right),
\end{equation}
as a function of $\mu/V$ for $J/V=0.14$. The thick solid lines represent the value of $\rho_{\bf Q}$ at the ground state. Points a (a') and b (b') correspond to the first-order transition (equal-energy) points. The curve of $\rho_{\bf Q}$ at the ground state exhibits finite jumps at these points as well as the curve of $\rho$ in Fig.~\ref{fig4}. The CMF method can calculate also metastable and unstable solutions.
The solution curve of $\rho_{\bf Q}$ for metastable (unstable) states is shown by the thin solid (dashed) lines. 
The total solution curves form one line of the SF solution ($e$-$f$) and one closed loop consisting of the solid and SS1 solutions ($a$-$b$-$y$-$x$-$a$). 
Such topology of the solution curves is in stark contrast to conventional first-order phase transitions, e.g., a typical liquid-solid transition. The formation of the separated solution curves originates from the re-entrance of the SF phase near the tip of $\rho=1/3$ (or $2/3$) solid lobes in the phase diagram upon sweeping the value of $\mu/V$. Of importance is that the ground-state SF solutions in the small and large $\mu/V$ sides are connected through the metastable solutions, which means that the SF state remains (at least locally) stable over the entire range of $\mu/V$ in Fig.~\ref{fig5}(a). This fact leads to a characteristic hysteresis behavior as described below.

Let us assume that we initially prepare a solid state which is located at point $i$ in Fig.~\ref{fig5}(a) as an initial state. When $\mu/V$ decreases, the system undergoes a melting transition to the SF phase though the transition path $i\rightarrow x\rightarrow x'\rightarrow e$ [upper panel, Fig.~\ref{fig5}(b)]. The solid state can remain metastable even after $\mu/V$ exceeds the first-order transition point $b$, and the melting of solid order occurs at the metastability limit of the SS1 phase (point $x$). 
Next, let us consider the reverse process (solidification) from the SF state at point $e$. In this case, the system remains in the SF phase upon increasing $\mu/V$ through $e\rightarrow f$ [lower panel, Fig.~\ref{fig5}(b)]. This is indeed attributable to the absence of the metastability limit of the SF phase. In a usual hysteresis, the reverse transition also can occur and the transition path forms a hysteresis loop. However, in this anomalous hysteresis, the transition can occur only unidirectionally from the solid to SF state, and the SF state cannot be solidified again by varying $\mu/V$. 
A similar anomalous hysteresis behavior also occurs when $\mu/V$ first increases and then decreases [see Fig.~\ref{fig5}(c)].

In practice, it is difficult to control the {\it global} chemical potential directly by external manipulation on cold atomic gases. However, one could vary the {\it local} chemical potential in the following way.
In actual experiments of ultracold gases, there exists a trap potential, e.g., $V_{t}({\bf r})=m\omega^2|{\bf r}|^2/2$ in addition to the optical lattice potential. 
Within the local-density approximation, the local chemical potential is written as $\tilde{\mu}_j=\mu-V_{t}({\bf r}_j)$. 
Therefore, we suggest that the anomalous hysteretic behavior could be realized by manipulating the frequency $\omega$ to control $\tilde{\mu}_j$ at the trap center.

\begin{figure}[t]
\includegraphics[scale=0.35]{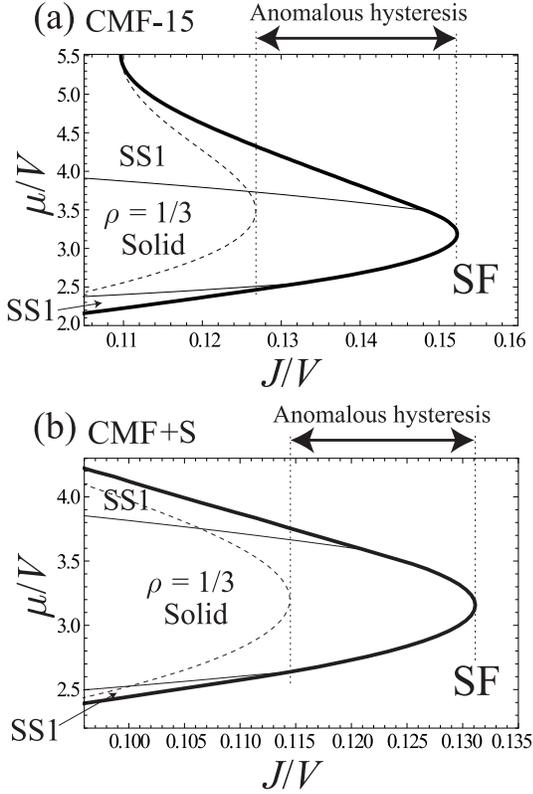}
\caption{\label{fig55}
The enlarged views of the phase diagrams obtained by the CMF-15 calculation (Fig.~\ref{fig1}) and the cluster-size scaling [Fig.~\ref{fig3}(b)]. The dashed lines indicate the metastability limit of the SF phase. In the region of the right of the dashed line}, the SF state is always (meta)stable. 
\end{figure}

Figures~\ref{fig55}(a) and \ref{fig55}(b) plot the metastability-limit line of the SF phase in the phase diagram obtained by CMF-15 and the cluster-size scaling. The anomalous hysteresis can be observed when the value of $J/V$ lies within the range sandwiched between the two vertical dotted lines, in which the SF state is always stable for any value of $\mu/V$. 
In this case, the lobe region consisting of the normal solid and SS1 phases is surrounded by the first-order transition boundary with the SF phase. 
In the presence of such a lobe region with the first-order transition boundary, the metastability limit of the surrounding phase (the SF phase in this case) is necessarily located inside the lobe as in Fig.~\ref{fig55}, and therefore there must be a finite region where the surrounding phase is always stable for any value of the parameter in the vertical axis ($\mu/V$ in this case). 
This implies that the anomalous hysteresis is not a unique phenomenon in this dipolar system but a common feature of the systems that have the geometry of the phase diagram mentioned above. 
In the following sections, in order to make this implication more convincing, we will provide a few other examples of the systems that exhibit the anomalous hysteresis.

\section{\label{3}Bose-Bose mixtures in a hypercubic lattice}
In this section, we consider the system described by the following two-component Bose-Hubbard model~\cite{jaksch-98}
\begin{eqnarray}
\hat{H}&=&
\sum_{\alpha=1,2}\left[-t_\alpha\sum_{\langle j,l \rangle}
(\hat{b}^{\dagger}_{j,\alpha} \hat{b}_{l,\alpha}+{\rm H.c.})-\mu_\alpha \sum_j \hat{n}_{j,\alpha}
\right.\nonumber\\
&&\left. +\frac{U_\alpha}{2}\sum_{ j}\hat{n}_{j,\alpha}(\hat{n}_{j,\alpha}-1)\right]+U_{12}\sum_{j}\hat{n}_{j,1}\hat{n}_{j,2},
\label{hamiltonian2}
\end{eqnarray}
where $\hat{b}^{\dagger}_{j,\alpha}$ creates a boson of the particle type $\alpha=1,2$ at site $j$ of a $D$-dimensional hypercubic lattice, $\hat{n}_{j,\alpha}=\hat{b}^{\dagger}_{j,\alpha}\hat{b}_{j,\alpha}$, $t_\alpha$ is the NN hopping amplitude, $U_\alpha$ the on-site intra-component interaction, $U_{12}$ the on-site inter-component interaction, and $\mu_\alpha$ the chemical potential. This model system could be realized using a mixture of two different types of bosons 
loaded into a sufficiently deep optical lattice~\cite{anderlini-07, widera-08, trotzky-08, weld-09, gadway-10, catani-08}. The strength of $U_{12}$ can be controlled experimentally by using Feshbach resonances~\cite{widera-08,thalhammer-08,tojo-10} or component-dependent optical lattices~\cite{gadway-10,mckay-10}.

In previous theoretical works, the ground-state phase diagram of Eq.~(\ref{hamiltonian2}) has been investigated for different parameter regions and dimensionality~\cite{altman-03, kuklov-04, chen-10, ozaki-12, chen-03, han-04, isacsson-05, mathey-07, arguelles-07, mathey-09, hu-09, hubener-09, sansone-10b, iskin-10}. In particular, it was found that the transition from SF to MI with even fillings can be of first order~\cite{kuklov-04, chen-10, ozaki-12}. However, all of them lacked the discussion about the metastability of the system and missed the possibility of the anomalous hysteresis. Since the MI region takes the lobe shape mentioned in the previous section, the anomalous hysteresis is expected to occur as in the triangular-lattice system of dipolar bosons. In the following we will show that this is indeed the case for both equal and slightly-unequal intra-component interactions. For observing the first-order SF-to-MI transition, the two-component system is advantageous over other systems in the sense that it has been already realized in several experiments~\cite{anderlini-07, widera-08, trotzky-08, weld-09, gadway-10, catani-08}. To stimulate further experimental efforts, we estimate the temperature and the lattice depth at which the first-order transitions can be clearly observed. Finally, we present the order-parameter theory for the first-order SF-to-MI transitions to explain the anomalous hysteresis within the Ginzburg-Landau framework.

\subsection{\label{3-1}Site-decoupling mean-field approximation}
We analyze the Hamiltonian in Eq.~(\ref{hamiltonian2}) using the simple (single-site) MF approximation~\cite{fisher-89, sheshadri-93, oosten-01, buonsante-04, lu-06}. Decoupling the hopping term as
\begin{eqnarray}
\hat{b}^{\dagger}_{j,\alpha} \hat{b}_{l,\alpha}&\approx& \langle \hat{b}^{\dagger}_{j,\alpha}\rangle  \hat{b}_{l,\alpha} +\langle \hat{b}_{l,\alpha}\rangle \hat{b}^{\dagger}_{j,\alpha} -\langle \hat{b}^{\dagger}_{j,\alpha}\rangle\langle \hat{b}_{l,\alpha}\rangle, \label{Decoupling}
\end{eqnarray}
we approximate the system by the sum of $M$ identical MF Hamiltonians: $\hat{H}\approx\sum_j\hat{H}_j^{\rm MF}$ with
\begin{eqnarray}
\hat{H}_j^{\rm MF}&=&
\sum_{\alpha=1,2}\left[-t_\alpha\sum_{\langle l\rangle}\phi_{l,\alpha}\left(\hat{b}_{j,\alpha}+\hat{b}_{j,\alpha}^\dagger-\phi_{j,\alpha}\right)-\mu_\alpha \hat{n}_{j,\alpha}
\right.\nonumber\\
&&\left. +\frac{U_\alpha}{2}\hat{n}_{j,\alpha}(\hat{n}_{j,\alpha}-1)\right]+U_{12}\hat{n}_{j,1}\hat{n}_{j,2}.
\label{lhamiltonian0}
\end{eqnarray}
Here the sum $\sum_{\langle l\rangle}$ runs over $z=2D$ nearest-neighbor sites of site $j$ and $M$ is the number of lattice sites.  The mean field $\phi_{j,\alpha}=\langle\hat{b}_{j,\alpha}\rangle=\langle\hat{b}_{j,\alpha}^\dagger\rangle$ plays the role of the SF order parameter. Here, we take the order parameters $\phi_{j,1}$ and $\phi_{j,2}$ to be real without loss of generality. 
Under the assumption of the spatial homogeneity $\phi_{j,\alpha}=\phi_{\alpha}$, the many-body lattice problem is reduced to a set of independent single-site problems with the effective Hamiltonian $\hat{H}_j^{{\rm MF}}$ and therefore we drop the site-labeling subscripts as
\begin{eqnarray}
\hat{H}_j^{{\rm MF}}&=&\hat{H}^{\rm MF}\nonumber\\
&=&\sum_{\alpha=1,2}\left[-zt_\alpha\phi_{\alpha}\left(\hat{b}_{\alpha}+\hat{b}_{\alpha}^\dagger-\phi_{\alpha}\right)-\mu_\alpha \hat{n}_{\alpha}
\right.\nonumber\\
&&\left. +\frac{U_\alpha}{2}\hat{n}_{\alpha}(\hat{n}_{\alpha}-1)\right]+U_{12}\hat{n}_{1}\hat{n}_{2}. 
\label{lhamiltonian2}
\end{eqnarray}
The value of the SF order parameter $\phi_\alpha$ is determined so that the free energy per site
\begin{eqnarray}
F(\phi_1,\phi_2)= -\frac{1}{\beta}{\rm Tr}\left(\exp [-\beta \hat{H}^{\rm MF}]\right) \label{FE}
\end{eqnarray}
takes a minimum value with respect to $\phi_1$ and $\phi_2$. Here $\beta=1/(k_{\rm B}T)$ is the inverse temperature. Around a first-order phase transition, the energy function has two or more local minima, one of which corresponds to the globally stable state and the others are metastable states. Additionally, $F(\phi_1,\phi_2)$ has several maxima or saddle points corresponding to unstable stationary states. Instead of the minimization of the free energy, we can also obtain $\phi_1$ and $\phi_2$ by directly solving the set of two self-consistent equations
\begin{eqnarray}
\phi_\alpha= {\rm Tr}\left(\hat{b}_{\alpha}  \exp [-\beta \hat{H}^{\rm MF}]\right)/{\rm Tr}\left(\exp[-\beta \hat{H}^{\rm MF}]\right) \label{SCEq}
\end{eqnarray}
with $\alpha=1$ and $2$.

It is known that the site-decoupling MF method can produce a qualitatively correct result for phase transitions between SF and MI phases~\cite{fisher-89,sheshadri-93}. This method becomes exact in the limit of infinite spatial dimensions. Therefore, the MF Hamiltonian $\hat{H}^{\rm MF}$ is expected to give a good approximation to binary Bose mixtures in a three-dimensional (simple-cubic) optical lattice with $z=6$. Another essential advantage for this study is that metastability analyses can be easily performed on the basis of the energy function $F(\phi_1,\phi_2)$.

\subsection{\label{3-2}Equal inter-component interactions}
\begin{figure}[t]
\includegraphics[scale=0.5]{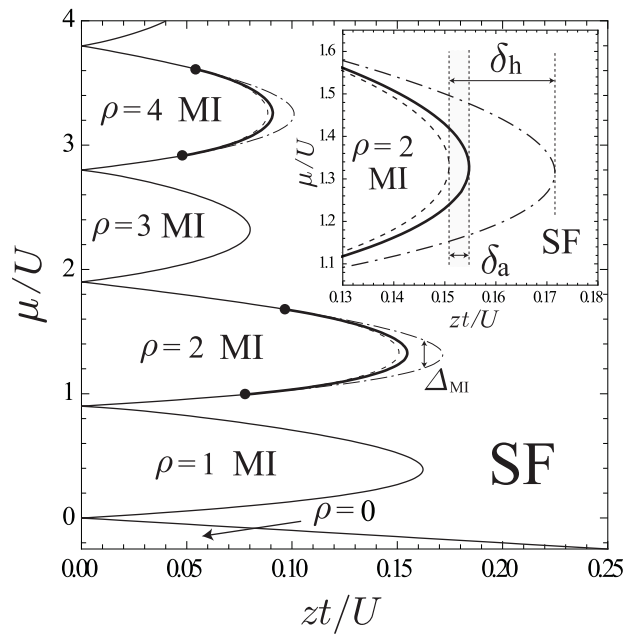}
\caption{\label{fig6}
The ground-state phase diagram of the two-component Bose-Hubbard model within the MF approximation for equal inter-component interactions and $U_{12}/U=0.9$. Second- and first-order phase transitions are indicated by the thin and thick lines, respectively. The phase boundaries are identical to those in Fig.~1(a) of Ref.~\cite{ozaki-12}. The dashed (dash-dotted) lines represent the metastability limits of SF (MI) phase, and the dots mark the tricritical points, where the transition changes from first to second order. The inset is an enlarged view of the region around the tip of the $\rho=2$ MI lobe. }
\end{figure}
In this section, we focus on the simple case of $t_1=t_2=t$, $U_1=U_2=U>0$, and $\mu_1=\mu_2=\mu$. Because of the $1 \leftrightarrow 2$ exchange symmetry of the Hamiltonian, we take $\phi_1=\phi_2=\phi$. 
We briefly review the MF ground-state phase diagram for $U_{12}/U=0.9$ shown in Fig.~\ref{fig6}, where the phase boundaries have been obtained in Fig.~1(a) of Ref.~\cite{ozaki-12}.
First-order phase transitions are found near the tips of the MI lobes (thick lines) unlike the SF-to-MI transition in the single-component Bose-Hubbard model.

\begin{figure}[t]
\includegraphics[scale=0.38]{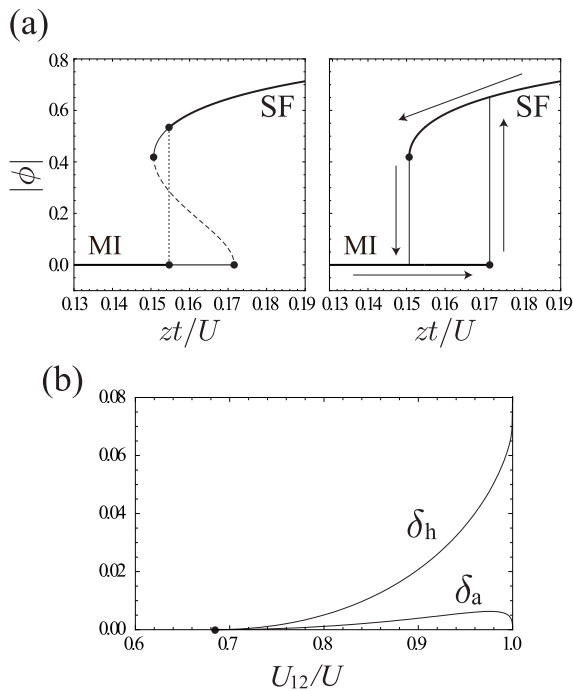}
\caption{\label{fig65}
(a) The MF solution curve (left) and the expected hysteresis loop (right) in the SF-to-MI transition for $\mu/U=1.314$ at zero temperature. In the left panel, the thick solid, thin solid, and dashed lines represent the ground, metastable, and unstable states, respectively. The vertical dotted line indicates the equal-energy point (the first-order transition point). (b) The $U_{12}/U$ dependence of $\delta_{\rm h}$ and $\delta_{\rm a}$. }
\end{figure}
We discuss the range of the hysteresis region and the temperature effects in the first-order transition between the SF and $\rho=2$ MI phases because they are essential for the visibility in actual experiments. We consider the SF-to-MI transition upon cycling (increasing and decreasing) $zt/U$. The corresponding experiment has been done for single-component Bose gases by tuning the depth of the lattice potential~\cite{trotzky-10, greiner-02, stoeferle-04, spielman-07}. In the case of Bose-Bose mixtures, the system is expected to show a hysteresis behavior in the SF-to-MI first-order transition. Figure~\ref{fig65}(a) shows the solution curve of Eq.~(\ref{SCEq}) and the expected hysteresis accompanying the phase transition between SF and MI $\rho=2$ for $\mu/U=1.314$. In this case, the transition process forms a conventional hysteresis loop (right panel). Unlike the anomalous hysteresis discussed in Sec.~\ref{2}, the transition occurs {\it bidirectionally} between the SF and MI phases.
The width $\delta_{\rm h}$ indicated by the arrow in the inset of Fig.~\ref{fig6} corresponds to the maximum size of the emergence region of the hysteresis behavior. We see that the hysteresis region is sufficiently sizable ($\approx 13\%$ of the size of the $\rho=2$ MI lobe in $zt/U$). 
As shown in Fig.~\ref{fig65}(b), the width $\delta_{\rm h}$ gets wider as $U_{12}/U$ is increased. If the inter-component repulsion is not sufficiently large ($U_{12}/U<0.68$), the hysteresis region vanishes and the SF-to-MI transition becomes second order at the entire boundary of the $\rho=2$ MI lobe~\cite{ozaki-12}.

\begin{figure}[t]
\includegraphics[scale=0.45]{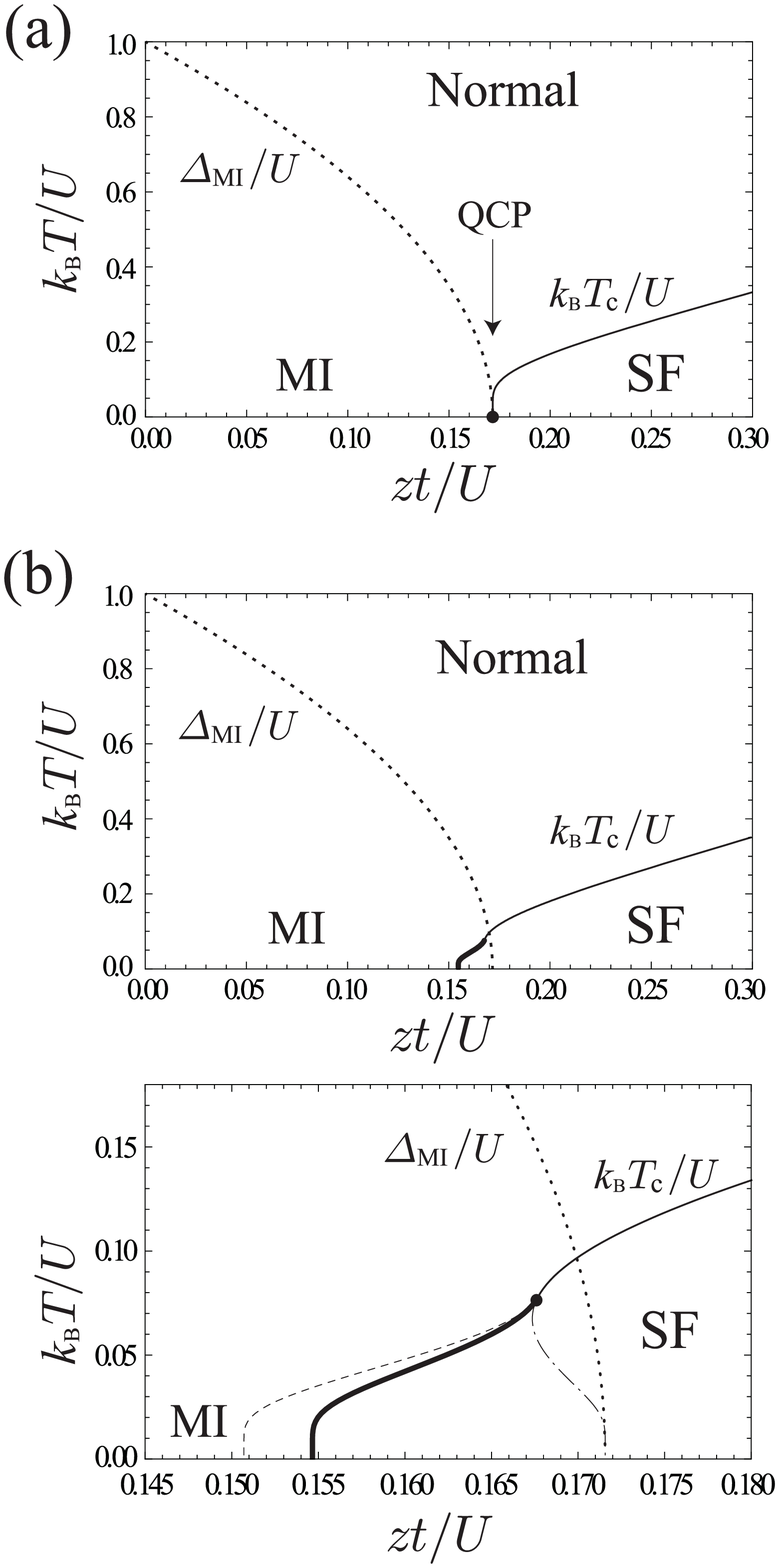}
\caption{\label{fig7}
Finite-temperature phase diagrams of the two-component Bose-Hubbard model for (a) $U_{12}/U=0$ and (b) $U_{12}/U=0.9$. The chemical potential $\mu/U$ is tuned to be the value at the tip of the $\rho=2$ MI lobe; $\mu/U=0.414$ for (a) and $\mu/U=1.314$ for (b). Second- and first-order phase transitions are indicated by the thin and thick lines, respectively. In addition to the SF transition temperature $T_{\rm c}/U$, we plot the value of the Mott gap $\Delta_{\rm MI}/U$ as a function of $zt/U$. Lower panel of (b): the enlarged view of the upper panel with adding the metastability limits of the SF (dashed line) and MI (dash-dotted line) phases. }
\end{figure}
\begin{figure}[t]
\includegraphics[scale=0.35]{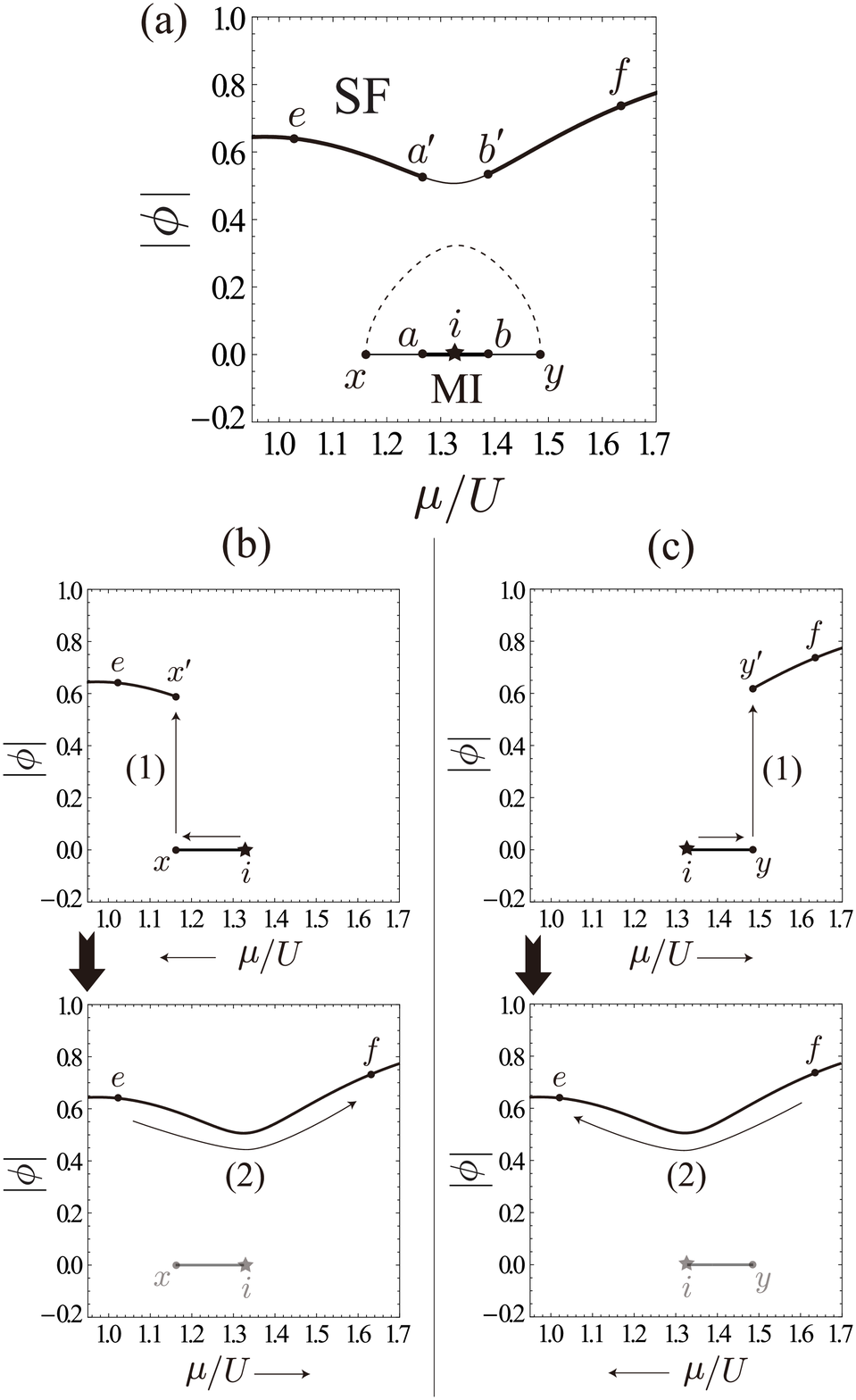}
\caption{\label{fig8}
(a) The solution curves for the SF order parameter $\phi$ as a function of $\mu/U$ at zero temperature and $zt/U=0.153$. The thick solid, thin solid, and dashed lines represent the ground, metastable, and unstable states, respectively. The anomalous hysteresis processes are shown in (b) and (c). The star indicates the position of the initial state for each case. The MI states represented by the gray line are ignored in the sweeping of $\mu/U$. }
\end{figure}

Let us discuss the effects of finite temperatures on the first-order nature of the SF-to-MI transition. 
Figure~\ref{fig7} shows the SF transition temperature $k_{\rm B}T_{\rm c}/U$ and the Mott gap $\Delta_{\rm MI}/U$ as functions of $zt/U$, where $U_{12}/U=0$ (a) and $0.9$ (b). 
When $U_{12}=0$, the Hamiltonian is completely separated into two independent single-component Bose systems. Therefore, the result shown in Fig.~\ref{fig7}(a) is identical to the SF-to-MI transition at unit filling in the single-component Bose-Hubbard model. At zero temperature, the system undergoes a continuous quantum phase transition from the SF to the MI phase at the critical ratio of the hopping amplitude to the interaction strength. The MF approximation leads to the critical value $t_{\rm c}/U=(3-2\sqrt{2})/z\approx 0.02860$ for a three-dimensional lattice ($z=6$)~\cite{oosten-01, lu-06}, which is somewhat smaller than the QMC result $t_{\rm c}/U=0.03409(2)$\cite{sansone-07}. In Fig.~\ref{fig7}, the chemical potential $\mu/U$ is fixed to be the value at the tip of the MI lobe, at which the transition occurs at a fixed integer filling. The effective action for the transition at this special multicritical point belongs to the $(D+1)$-dimensional $XY$ universality class~\cite{fisher-89}. Above the upper critical dimension ($D>3$), the critical temperature $T_{\rm c}$ for the SF-to-normal transition decreases according to the following scaling in the vicinity of the quantum critical point (QCP)~\cite{sachdev-97}:
\begin{eqnarray}
T_{\rm c}\sim (t-t_{\rm c})^{z_{\rm c}\nu/(1+\theta \nu)}
\end{eqnarray}
with the dynamical critical exponent $z_{\rm c}=1$, the correlation length exponent $\nu=1/2$, and the scaling dimension $\theta =D+z_{\rm c}-4$. 
The site-decoupling MF (or Gutzwiller) approximation becomes exact in the limit of infinite dimensions. Therefore, the curve of $k_{\rm B}T_{\rm c}/U$ in Fig.~\ref{fig7}(a) exhibits a precipitous fall at the QCP according to $T_{\rm c}\sim (t-t_{\rm c})^{1/(D-1)}$ with $D\rightarrow \infty$.
The transition temperature $T_{\rm c}$ and the Mott gap $\Delta_{\rm MI}$ simultaneously vanish at the QCP~\cite{trotzky-10}. Notice that the Mott gap $\Delta_{\rm MI}$ corresponds to the width of the MI lobe in $\mu$ at zero temperature. In a strict sense, the MI phase exists only at $T=0$. However, for low temperatures, Mott-like features still remain in the normal fluids, where the filling factor deviates only slightly from integer values.

When $U_{12}/U>0.68$, the situation drastically differs from the single-component (or $U_{12}/U=0$) case. As shown in Fig.~\ref{fig7}(b), the SF-to-MI transition for low temperatures becomes first-order and thus the QCP vanishes. In the vicinity of the first-order transition, $\Delta_{\rm MI}$ is given by the width between the upper and lower metastability limits of the MI phase (see the phase diagram in Fig.~\ref{fig6}). The first-order nature of the SF-to-MI (or SF-to-normal) transition remains until $T\approx 0.08U$ for $U_{12}/U=0.9$.

Next we consider variations of $\mu/U$ at fixed $zt/U$ such that the anomalous hysteresis occurs.
We see in Fig.~\ref{fig6} that the system undergoes a re-entrant first-order transition from SF to MI, and back to SF 
near the tip of the Mott lobe with even fillings. For a certain value of $zt/U$, the SF phase remains stable for any $\mu/U$, and the hysteresis process shows the anomalous unidirectional behavior. In Fig.~\ref{fig8}(a), we show the solution curves of Eq.~(\ref{SCEq}) as a function of $\mu/U$ near the transition between the SF and the $\rho=2$ MI phase for $zt/U=0.153$ at zero temperature. Notice that the solution curves are completely separated into one closed loop and one line unlike a conventional first-order transition [compare with Fig.~\ref{fig65}(a)]. Thus, the hysteresis exhibits a unidirectional behavior upon cycling $\mu/U$.

\begin{figure}[t]
\includegraphics[scale=0.45]{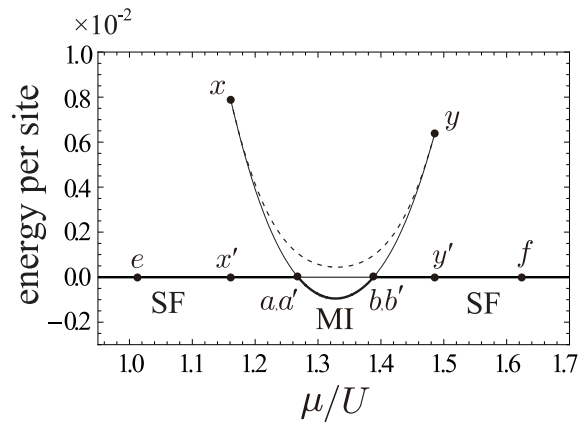}
\caption{\label{fig9}
The curve of the MF energy relative to the energy of the stable SF state in the re-entrant first-order transition between SF and MI. The thick solid, thin solid, and dashed lines are the global minimum, local minimum, and maximum values of the energy landscape, respectively. Each point corresponds to the point with the same letter in Fig.~\ref{fig8}. }
\end{figure}

The transition process is similar to the one of the triangular-lattice dipolar system discussed in Sec.~\ref{2}. Starting from an initial MI state at point $i$, if $\mu/U$ is decreased and then increased, the transition from MI to SF can occur ($i\rightarrow x\rightarrow x'\rightarrow e$). However, the system does not return to MI ($e\rightarrow f$) as shown in Fig.~\ref{fig8}(b). The unidirectional behavior is also obtained upon first increasing and then decreasing $\mu/U$ [Fig.~\ref{fig8}(c)]. The energy curve in the anomalous hysteresis has a characteristic shape where a line crosses a closed loop as shown in Fig.~\ref{fig9}. This is different from the case of a conventional first-order transition, where the hysteresis has a so-called swallowtail structure~\cite{swallowtail}. The anomalous behavior can be seen when $zt/U$ lies within the width $\delta_{\rm a}$ in the inset of Fig.~\ref{fig6}, where the SF phase is stable for any $\mu/U$. As shown in Fig.~\ref{fig65}(b), the width $\delta_{\rm a}$ exhibits a maximum at $U_{12}/U\approx 0.98$.

\subsection{\label{3-3}Unequal intra-component interactions}
While we assumed $U_1=U_2$ for theoretical simplicity in Sec.~\ref{3-2}, this condition does not hold strictly in actual experimental systems of Bose-Bose mixtures. We consider here the case of $U_1\approx U_2$ to discuss the first-order SF-to-MI transitions in a realistic situation. Specifically, we suppose a binary mixture of $^{87}$Rb atoms in the two hyperfine states, $|F=1,m_F=1\rangle$ and $|F=2,m_F=-1\rangle$, which has been created in several previous experiments~\cite{anderlini-07, widera-08, trotzky-08, weld-09, gadway-10}. We label the former (latter) hyperfine state as $\alpha=1$ ($\alpha=2$). Since the scattering lengths for the intra-component interactions are given by $a_{1}=100.40a_{\rm B}$ and $a_{2}=95.00a_{\rm B}$~\cite{mertes-07,tojo-10}, $U_1$ and $U_2$ are slightly unequal as $U_2/U_1=0.9462$, where $a_{\rm B}$ is the Bohr radius. We recall that $U_{12}$ can be widely controlled using the Feshbach resonance or the component-dependent optical lattice, while the bare scattering length is given by $a_{12}=97.66 a_{\rm B}$. When $U_1\neq U_2$,
the SF order parameters $\phi_1$ and $\phi_2$ can take different values, and it is even possible that one component of bosons forms a MI state while another component is condensed; e.g., $\phi_1=0$ and $\phi_2\neq 0$~\cite{kuklov-04}. We call this phase MI$_1$+SF$_2$ at zero temperature and NF$_1$+SF$_2$ at finite temperatures, where NF means a normal fluid.
We tune the difference of the chemical potentials $\mu_1-\mu_2$ so that the populations of the two components are balanced, i.e., $\langle \hat{n}_1\rangle=\langle \hat{n}_2\rangle$. The total filling factor $\rho$ is controlled by the averaged chemical potential $\bar{\mu}\equiv \sum_{\alpha}\mu_\alpha/2$.

\begin{figure}[b]
\includegraphics[scale=0.5]{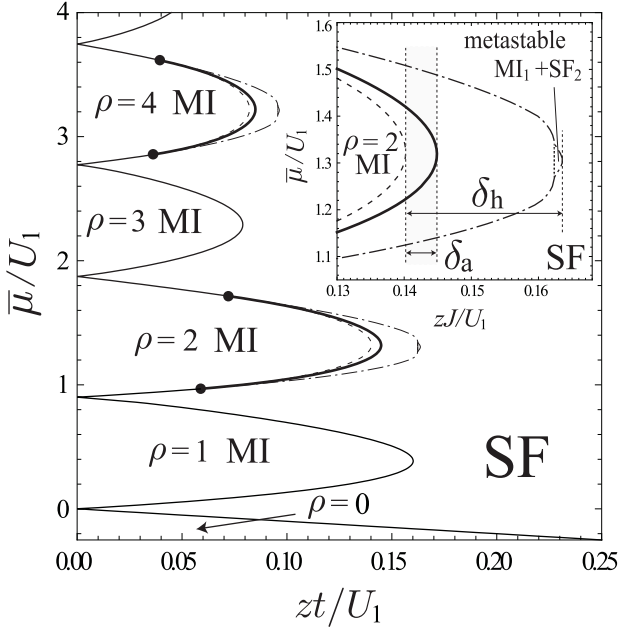}
\caption{\label{fig91}
The ground-state phase diagram of the two-component Bose-Hubbard model within the MF approximation for $U_{2}/U_1=0.9462$ and $U_{12}/U_1=0.9$. Second- and first-order phase transitions are indicated by the thin and thick lines, respectively. The dashed (dash-dotted) lines represent the metastability limits of SF (MI) phase, and the dots mark the tricritical points, where the transition changes from first to second order. The inset is an enlarged view of the region around the tip of the $\rho=2$ MI lobe. } 
\end{figure}
\begin{figure}[t]
\includegraphics[scale=0.45]{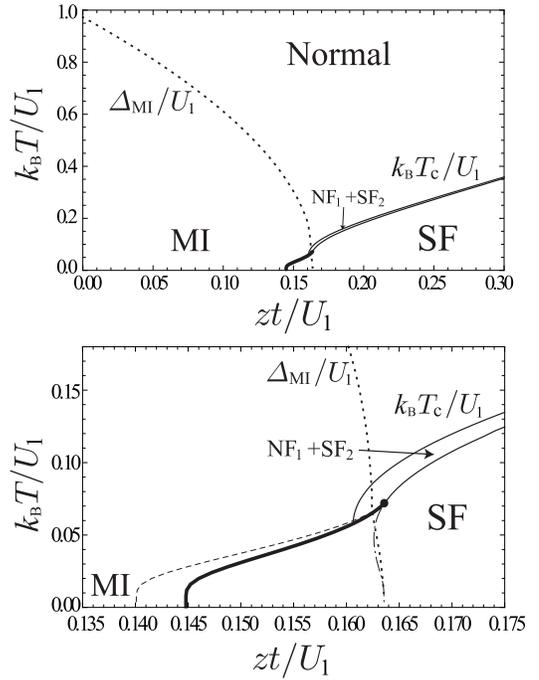}
\caption{\label{fig92}
Finite-temperature phase diagram for $U_{2}/U_1=0.9462$ and $U_{12}/U_1=0.9$. The chemical potential $\mu/U_1$ is fixed to be $1.305$. Second- and first-order phase transitions are indicated by the thin and thick lines, respectively. We also plot the Mott gap $\Delta_{\rm MI}/U_1$ for the $\alpha=1$ bosons (dotted lines). Lower panel: the enlarged view of the upper panel with adding the metastability limits of the SF (dashed line) and MI (dash-dotted line) phases.} 
\end{figure}

Solving the set of Eqs.~(\ref{SCEq}) for $\phi_1$ and $\phi_2$, we determine the ground-state phase diagram for $t_1=t_2=t$, $U_2/U_1=0.9462$, and $U_{12}/U_1 = 0.9$. We see in Fig.~\ref{fig91} that the slight difference of the intra-component interactions hardly changes the phase diagram. The MI$_1$+SF$_2$ phase emerge only as a metastable state in a very small region near the tip of the metastability limit of each MI phase (see the inset). In addition, Fig.~\ref{fig92} shows that the finite-temperature phase diagram also remains almost the same as the case of $U_1=U_2$ in Fig.~\ref{fig7}(b). The only noticeable change is the emergence of a narrow region of the NF$_1$+SF$_2$ phase at finite temperatures. 

Since there remain the first-order SF-to-MI transitions and the hysteresis, it is expected that one can simulate them by using the binary mixture of $^{87}$Rb atoms in optical lattices~\cite{anderlini-07, widera-08, trotzky-08, weld-09, gadway-10}. As seen in Fig.~\ref{fig92}, the temperature range where the first-order nature of the SF-to-MI transition could be clearly observed is given by $k_{\rm B}T< 0.02U_1$. The hysteresis loop at a fixed chemical potential occurs in the range of $0.14< zt/U_1< 0.163$ ($\delta_{\rm h}$ in the inset of Fig.~\ref{fig91}), while the anomalous hysteresis behavior upon cycling the value of $\mu/U_1$ is expected to occur when the ratio $zt/U$ lies within $0.14< zt/U_1< 0.145$ ($\delta_{\rm a}$). 
In experiments with optical lattices, the ratio $t/U_1$ can be tuned by manipulating the maximum potential depth $V_0$. To connect the experimental parameters to the Bose-Hubbard parameters, we use the following formulae~\cite{bloch-08},
\begin{eqnarray}
J= \frac{4}{\sqrt{\pi}}E_{r}\left(\frac{V_0}{E_r}\right)^{3/4}\exp \left[-2\left(\frac{V_0}{E_r}\right)^{1/2}\right]
\end{eqnarray}
and 
\begin{eqnarray}
U=\sqrt{\frac{8}{\pi}}kaE_{r}\left(\frac{V_0}{E_r}\right)^{3/4},
\end{eqnarray}
where $E_r$ is the recoil energy. Assuming a simple-cubic optical lattice with a lattice constant $d=\pi/k=532$ nm and the scattering length $a=a_{1}=100.40a_{\rm B}$, we obtain the required tuning ranges for observing the hysteresis loop as $13.7 < V_0/E_r < 14.3$. This level of controllability of $V_0/E_r$ can be achieved by current techniques~\cite{trotzky-10}, in which $V_0/E_r$ can be tuned at least in 0.25 increments. To observe the anomalous hysteresis, the lattice depth has to be fine tuned to be within the range of $14.2 < V_0/E_r < 14.3$, which is a challenge for future experimental efforts. In addition, the system needs to be cooled down to $T\lesssim 0.02U_1\approx 0.70$ nK for a clear observation of the hysteresis behavior. This is much higher than the temperature required to observe magnetism in optical-lattice systems and expected to be achieved in the near future, given that the currently accessible temperatures are $\approx 1$ nK~\cite{weld-09}. 

\subsection{\label{3-4}The Ginzburg-Landau description of the anomalous hysteresis}
In the previous sections, we have seen the anomalous hysteresis behavior in two systems. Theoretically, the two-component Bose-Hubbard system is much simpler to handle compared to the dipolar model in Sec.~\ref{2} because the anomalous hysteresis can be described only at the single-site MF level with one order parameter $\phi$. Taking advantage of this simplicity, we construct in this section the Ginzburg-Landau theory for the anomalous hysteresis by taking the two-component Bose-Hubbard model with $U_1=U_2$ as an example. Using this approach, we will clarify the reason why the hysteresis shows the anomalous unidirectional behavior.

We construct the Ginzburg-Landau energy function by using the perturbative MF method~\cite{oosten-01,tsuchiya-04,mitra-08} in order to express the coefficients in the energy function as functions of the microscopic parameters in the Hamiltonian. Here, we assume $t_1=t_2=t$, $U_1=U_2=U>0$, and $\mu_1=\mu_2=\mu$, and take the SF order parameter $\phi_1=\phi_2=\phi$ to be real without loss of generality. We divide the MF Hamiltonian $\hat{H}^{\rm MF}$ into diagonal and off-diagonal parts as 
$\hat{H}^{\rm MF}=\hat{H}_{0}+\hat{V}+2zt\phi^2$, where
\begin{eqnarray}
\hat{H}_{0}=\sum_{\alpha}\left[-\mu\hat{n}_{\alpha}+\frac{U}{2}\hat{n}_{\alpha}(\hat{n}_{\alpha}-1)\right]+U_{12}\hat{n}_{1}\hat{n}_{2}
\label{hamiltonianU}
\end{eqnarray}
and
\begin{eqnarray}
\hat{V}=-zt\phi\sum_{\alpha}(\hat{b}_{\alpha}+\hat{b}_{\alpha}^\dagger).
\label{hamiltoniant}
\end{eqnarray}
Assuming that $\phi$ is small, we deal with $\hat{V}$ as a small perturbation. The eigenvalue of the unperturbed Hamiltonian $\hat{H}_{0}$ for the two-component Fock state $|n_1,n_2\rangle$ is given as
\begin{eqnarray}
\hat{H}_{0}|n_1,n_2\rangle=E_0(n_1,n_2)|n_1,n_2\rangle,
\end{eqnarray}
where
\begin{eqnarray}
E_0(n_1,n_2)&=&-\mu (n_1+n_2)+\frac{U}{2}n_1(n_1-1)\nonumber\\
&&+\frac{U}{2}n_2(n_2-1)+U_{12}n_1n_2.
\end{eqnarray}
Now we focus on the first-order phase transitions which appear near the tips of MI lobes with even fillings $\rho\equiv 2m$ ($m=0,1,2,\dots$). For even fillings, the ground state of $\hat{H}_{0}$ is simply given by $|{\rm i}\rangle\equiv |m,m\rangle$. 
The second-order correction to the energy is given by
\begin{eqnarray}
\delta E_2&=&\langle {\rm i}|\hat{V}_\circ \hat{G}_\circ\hat{V}|{\rm i}\rangle+2zt\phi^2\nonumber\\
&=&\sum_{{\rm p}}^\prime\frac{|\langle {\rm i}|\hat{V}|{\rm p}\rangle|^2}{E_{\rm i}-E_{\rm p}}+2zt\phi^2,
\end{eqnarray}
where we used the notations $\hat{G}=(E_{\rm i}-\hat{H}_{0})^{-1}$ and $\hat{A}_\circ\hat{B}=\hat{A}\sum_{\rm p}^\prime|{\rm p}\rangle \langle {\rm p}|\hat{B}$. Here the sum $\sum_{{\rm p}}^\prime$ runs over all the eigenstates of $\hat{H}^{0}$ other than the initial state $|{\rm i}\rangle= |m,m\rangle$. $E_{\rm p}=E_0(n_1,n_2)$ is the eigenvalue for the state $|{\rm p}\rangle=|n_1,n_2\rangle$. Note that the last term comes from the MF decoupling in Eq.~(\ref{Decoupling}). In the second-order perturbation process, we have only four intermediate states $|{\rm p}\rangle=|m\pm 1,m\rangle$ and $|m,m\pm 1\rangle$, and the matrix elements $\langle {\rm i}|\hat{V}|{\rm p}\rangle$ are easily calculated. 
As a result, we obtain
\begin{eqnarray}
\delta E_2&=&\Bigg[\frac{2(zt)^2m}{-\mu+U(m+1)+U_{12}m}\nonumber\\
&+&\frac{2(zt)^2(m+1)}{\mu-(U+U_{12})m}+2zt\Bigg]\phi^2\nonumber\\
&\equiv& a_2\phi^2, \label{E2}
\end{eqnarray}
where $a_2=a_2(zt,U,U_{12},\mu,m)$ is the second-order coefficient of the Ginzburg-Landau energy function. In order to discuss the first-order transition phenomena in this system, one needs to consider the terms up to the sixth-order in $\phi$. The fourth-order and sixth-order corrections to the energy are given, respectively, by
\begin{eqnarray}
\delta E_4&=&\langle {\rm i}|\hat{V}_\circ \hat{G}_\circ\hat{V}_\circ \hat{G}_\circ\hat{V}_\circ \hat{G}_\circ\hat{V}|{\rm i}\rangle-\delta \bar{E_2}\langle {\rm i}|\hat{V}_\circ \hat{G}^2_\circ\hat{V}|{\rm i}\rangle\nonumber\\
&\equiv& a_4\phi^4 \label{E4}
\end{eqnarray}
and
\begin{eqnarray}
\delta E_6&=&\langle {\rm i}|\hat{V}_\circ \hat{G}_\circ\hat{V}_\circ \hat{G}_\circ\hat{V}_\circ \hat{G}_\circ\hat{V}_\circ \hat{G}_\circ\hat{V}_\circ \hat{G}_\circ\hat{V}|{\rm i}\rangle\nonumber\\
&&-\delta E_4\langle {\rm i}|\hat{V}_\circ \hat{G}^2_\circ\hat{V}|{\rm i}\rangle+(\delta \bar{E_2})^2\langle {\rm i}|\hat{V}_\circ \hat{G}^3_\circ\hat{V}|{\rm i}\rangle\nonumber\\
&&+\delta \bar{E_2}\langle {\rm i}|\hat{V}_\circ \hat{G}^2_\circ\hat{V}_\circ \hat{G}_\circ\hat{V}_\circ \hat{G}_\circ\hat{V}|{\rm i}\rangle\nonumber\\
&&+\delta \bar{E_2}\langle {\rm i}|\hat{V}_\circ \hat{G}_\circ\hat{V}_\circ \hat{G}^2_\circ\hat{V}_\circ \hat{G}_\circ\hat{V}|{\rm i}\rangle\nonumber\\
&&+\delta \bar{E_2}\langle {\rm i}|\hat{V}_\circ \hat{G}_\circ\hat{V}_\circ \hat{G}_\circ\hat{V}_\circ \hat{G}^2_\circ\hat{V}|{\rm i}\rangle\nonumber\\
&\equiv& a_6\phi^6. \label{E6}
\end{eqnarray}
Here, $\delta \bar{E_2}=\delta E_2-2zt\phi^2$. 
One has to consider up to $16 \times 8$ matrix elements of $\hat{V}$ between intermediate states in the calculation of the sixth-order coefficient $a_6=a_6(zt,U,U_{12},\mu,m)$, since many intermediate states emerge. However, for $\rho=2$ ($m=1$), the maximum size of the matrix reduces to $10\times 6$ due to the lower limit of occupation ($n_1,n_2 \geq 0$).

\begin{figure}[t]
\includegraphics[scale=0.34]{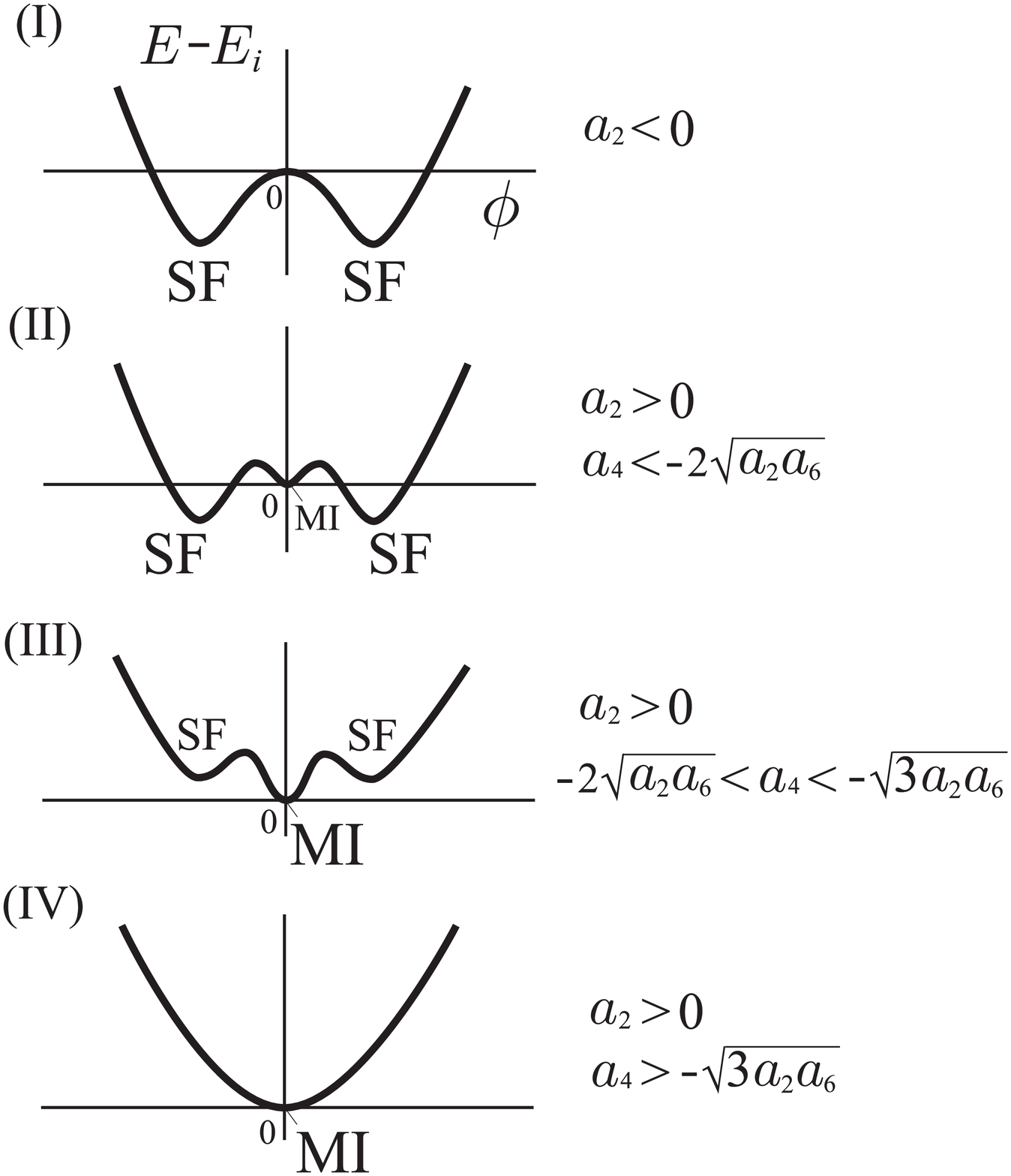}
\caption{\label{fig10}
Schematic illustrations of energy landscapes. Here, the order parameter $\phi$ is taken to be real. 
}
\end{figure}
Now we obtain the Ginzburg-Landau energy
\begin{eqnarray}
E(\phi)=E_{\rm i}+a_2\phi^2+a_4\phi^4+a_6\phi^6, \label{EGL}
\end{eqnarray}
whose coefficients are related to the microscopic system parameters in the two-component Bose-Hubbard Hamiltonian through Eqs.~(\ref{E2}-\ref{E6}). The shape of the energy as a function of $\phi$ changes depending on the values of the coefficients. For the existence of a stable ground state, at least the highest-order coefficient $a_6$ must be positive. If the second-order coefficient $a_2$ is negative, the energy function forms two minima (the wine-bottle or Mexican-hat shape when the imaginary axis of $\phi$ is considered) shown in Fig.~\ref{fig10}(I). Therefore, the ground state is a SF state with a finite $\phi$. Although the MI state with $\phi=0$ is also a stationary point ($dE(\phi)/d\phi=0$), it is unstable. If $a_2$ becomes positive, the profile around $\phi=0$ changes into a convex shape, and the MI state becomes stable. If the transition is of second order, the condition $a_2=0$ gives the phase boundary between SF and MI. However, in first-order transitions, the energy can have three minima, corresponding to a MI state and two equivalent SF states, shown in Figs.~\ref{fig10}(II) and~\ref{fig10}(III). There are two maxima which correspond to unstable SF states. The energy of the SF states at the minima is easily obtained by substituting the solution of $dE(\phi)/d\phi=0$ into Eq.~(\ref{EGL}). If $a_4<-2\sqrt{a_2a_6}$, the energy of the SF state is smaller than that of the MI state, and the energy function acquires the shape shown in Fig.~\ref{fig10}(II). In contrast, the MI state has the lowest energy for $a_4>-2\sqrt{a_2a_6}$ as shown in Fig.~\ref{fig10}(III). When the value of $a_4$ exceeds $-\sqrt{3a_2a_6}$, the stationary solutions with $\phi\neq 0$ disappear and the energy profile has only one minimum at $\phi = 0$ as shown in Fig.~\ref{fig10}(IV). 
In summary, the expansion up to sixth order in $\phi$ can describe four kinds of energy landscapes, which are separated by the conditions $a_2=0$, $a_4^2=4a_2a_6$, and $a_4^2=3a_2a_6$. These conditions correspond to the metastability limit of MI, the first-order transition boundary (or the equal-energy point), and the metastability limit of SF, respectively.

\begin{figure}[t]
\includegraphics[scale=0.4]{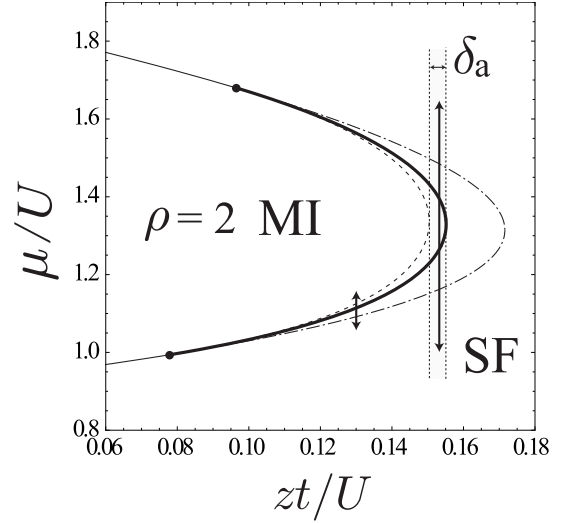}
\caption{\label{fig11}
The phase boundary of the transition between the SF and $\rho=2$ MI phases obtained by the perturbative MF method. We set $U_{12}/U=0.9$ as in Fig.~\ref{fig6}. The dashed (dash-dotted) lines represent the metastability limit of SF (MI) phase, and the vertical lines with arrowheads mark the plot ranges of Figs.~\ref{fig12} and~\ref{fig13}. In the shaded region with width $\delta_{\rm a}$, the anomalous hysteresis appears when varying $\mu/U$. The tricritical points (dots) can be obtained from the condition $a_2=a_4=0$.
}
\end{figure}
The coefficients $a_2$, $a_4$, and $a_6$ are related to $zt$, $U$, $U_{12}$, and $\mu$ through the perturbative MF method. Therefore, the phase diagram in the ($zt/U$,$\mu/U$) plane is easily obtained by the three conditions $a_2=0$, $a_4^2=4a_2a_6$, and $a_4^2=3a_2a_6$. In Fig.~\ref{fig11}, we show the curves of the first-order transition boundary between the SF and $\rho=2$ MI phases and the metastability limits of MI and SF for $U_{12}/U=0.9$. Compared to the full MF result in Fig.~\ref{fig6}, while the same result is reproduced for the metastability limit of MI, the positions of the first-order transition boundary and the metastability limit of SF are slightly shifted. This is because higher-order terms are neglected in the perturbative expansion of the energy function.

\begin{figure}[t]
\includegraphics[scale=0.37]{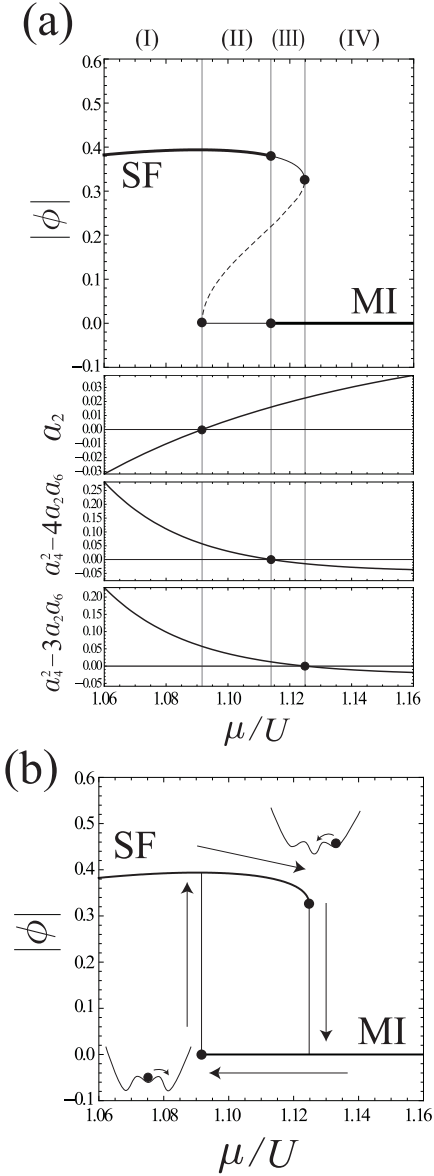}
\caption{\label{fig12}
(a) The solution curve for the order parameter $\phi$ as a function of $\mu/U$ obtained by the perturbative MF method. We set $zt/U=0.13$. The thick solid, thin solid, and dashed lines represent the ground, metastable, and unstable states, respectively. In the regions I to IV, the energy function has the shape shown in Fig.~\ref{fig10}, respectively. (b) The corresponding hysteresis loop structure when cycling the value of $\mu/U$. 
}
\end{figure}
Let us see the changes in the shape of the energy function and the behaviors of the coefficients during the anomalous hysteresis induced by cycling the chemical potential $\mu/U$. First, we review the case of the conventional hysteresis loop. The value of the SF order parameter $\phi$ is obtained as
\begin{eqnarray}
\phi^2=\frac{-a_4 + \sqrt{a_4^2 - 3 a_2 a_6}}{3 a_6},~\frac{-a_4 - \sqrt{a_4^2 - 3 a_2 a_6}}{3 a_6}\label{phi}
\end{eqnarray}
from $d E(\phi)/d\phi=0$. The latter solution corresponds to the unstable SF states at the maxima of the energy function, which exist only for $a_4<0$ and $a_4^2>3a_2a_6$, i.e., in the case of Figs.~\ref{fig10}(II) and~\ref{fig10}(III). We plot the solution curve as a function of $\mu/U$ for $zt/U=0.13$ in Fig.~\ref{fig12}(a). The lower panels show the coefficients of the Ginzburg-Landau energy as functions of $\mu/U$. The three conditions $a_2=0$, $a_4^2=4a_2a_6$, and $a_4^2=3a_2a_6$ separate the plot into four regions. The numbers I to IV indicate that the energy profile has the shape shown in Fig.~\ref{fig10}. In this case, the shape of the energy function changes from (I) to (IV) as the value of $\mu/U$ increases. One finds three stationary solutions (stable MI and SF solutions and an unstable SF solution) in the intermediate regions (II) and (III).

The ground state of regions (I) and (II) is a SF state, which is located at the global minima of $E(\phi)$. If one increases $\mu/U$ from the initial SF state, the system remains in the metastable SF phase even in region (III) since $E(\phi)$ still has local minima at finite $\phi$. The SF state is destabilized in region (IV) and the transition to the MI phase occurs due to the disappearance of the local minima [see the illustrations in Fig.~\ref{fig12}(b)]. On the other hand, when starting from an initial MI state in region (IV), it remains locally stable until region (II) and then the system undergoes a transition to the SF phase in region (I). As a result, the hysteresis cycle forms a loop structure as shown in Fig.~\ref{fig12}(b).

\begin{figure}[t]
\includegraphics[scale=0.37]{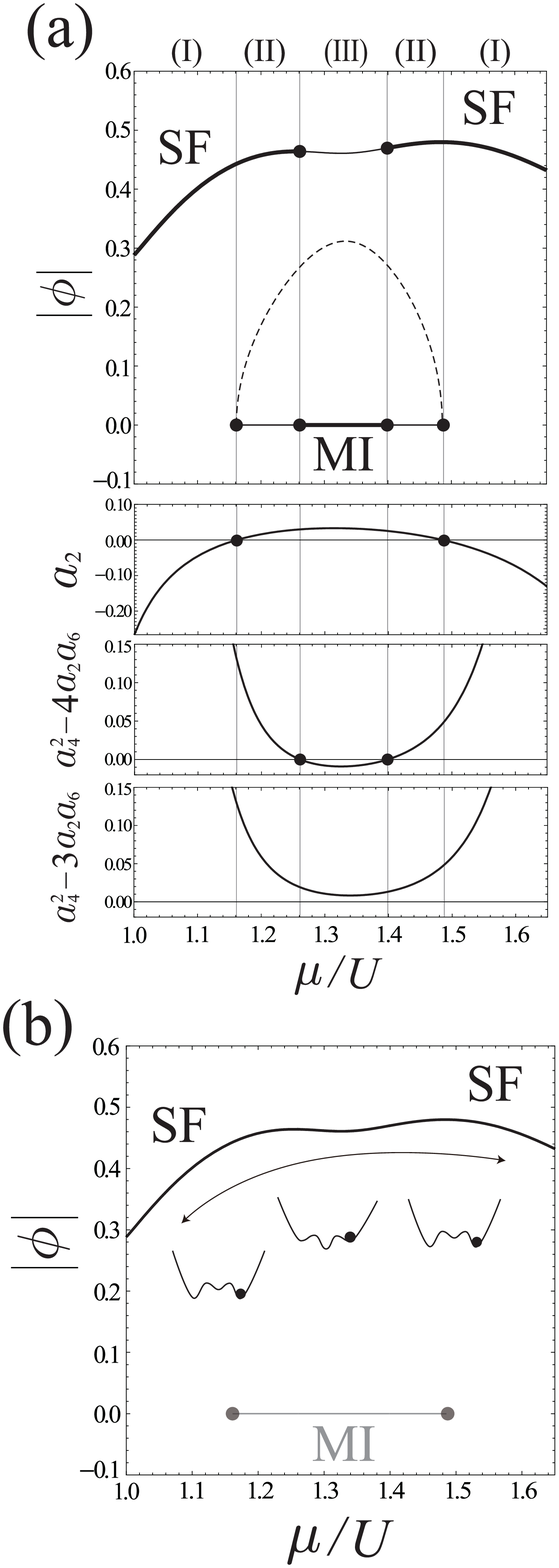}
\caption{\label{fig13}
(a) The solution curves for the SF order parameter $\phi$ as a function of $\mu/U$ obtained by the perturbative MF method. We set $zt/U=0.153$. The thick solid, thin solid, and dashed lines represent the ground, metastable, and unstable states, respectively. (b) The anomalous behavior in the hysteresis process. The system keeps in the SF phase ignoring the solutions of MI states represented by the gray line.
}
\end{figure}
For $zt/U$ close to the tip of the MI lobe, the hysteresis process exhibits the anomalous behavior discussed in Sec.~\ref{3-1}. The upper panel of Fig.~\ref{fig13}(a) shows the solution curve of $\phi$ as a function of $\mu/U$ for $zt/U=0.153$. The full MF result shown in Fig.~\ref{fig8}(a) can be reproduced almost perfectly by the perturbative MF calculation with up to the sixth-order perturbations. In this case, the system can undergo the transition only unidirectionally from the MI to SF phase as shown in Figs.~\ref{fig8}(b) and~\ref{fig8}(c). The cause of this anomalous hysteresis can be understood from the behaviors of the coefficients in the Ginzburg-Landau energy plotted in the lower panels of Fig.~\ref{fig13}(a). Unlike in Fig.~\ref{fig12}(a), the curve of $a_4^2-4a_2a_6$ in Fig.~\ref{fig13}(a) crosses $0$ twice since the energies of SF and MI become identical at the two points. The curve of $a_2$ also crosses $0$ twice at the points which correspond to the metastability limits of the MI phase. However, the curve of $a_4^2-3a_2a_6$ does {\bf not} cross $0$, which means that the coefficients always satisfy the inequality $a_4<-\sqrt{3a_2a_6}$ in the anomalous hysteresis. This fact is essential for the emergence of the anomalous hysteresis. The behaviors of the coefficients indicate that the energy profile changes as (I)$\rightarrow$(II)$\rightarrow$(III)$\rightarrow$(II)$\rightarrow$(I) without going through region (IV) when $\mu/U$ increases. Therefore, $E(\phi)$ always has minima at $\phi\neq 0$ and a stable SF state exists over the entire range of $\mu/U$. Consequently, the system is caught in a local minimum of $E(\phi)$ at $\phi\neq 0$ and the SF state is not destabilized dynamically [see illustrations in Fig.~\ref{fig13}(b)]. This is the cause of the unidirectional transition process in the anomalous hysteresis.

When $U_1 \neq U_2$, the sixth-order Ginzburg-Landau energy is given as a function of $\phi_1$ and $\phi_2$. The energy function $E(\phi_1,\phi_2)$ consists of many terms proportional to  $\propto \phi_1^{2m}\phi_2^{2n}$ ($0\leq m+n\leq 3$). When $|U_1 - U_2| \ll U_1$ (e.g., in the case of Sec.~\ref{3-2}), the values of $\phi_1$ and $\phi_2$ are almost identical 
and the first-order SF-to-MI transitions should be also described by the sixth-order Ginzburg-Landau energy in a similar way.

\section{\label{4}Spin-1 Bosons}
In Sec.~\ref{3-4}, we have shown that the anomalous hysteresis can be qualitatively understood from the Ginzburg-Landau theory, in which the dependence of the coefficients on microscopic parameters of the system plays an essential role in causing the unidirectional behavior. The anomalous hysteresis is now mathematically well established as another type of first-order transition phenomena different from the conventional hysteresis-loop formation. However, such an anomalous hysteresis has never been observed in experiments on either cold-atom systems or solid-state materials. Therefore, in this section we add another example to the list of possible systems that exhibit anomalous hysteresis in order to provide further guidance in achieving the first experimental observation.

Here, we consider the system of spin-1 bosons confined to an optical lattice~\cite{krutitsky-04,krutitsky-05,kimura-05,kimura-06,pai-08,tsuchiya-04,rizzi-05,bergkvist-06,batrouni-09}. In previous studies~\cite{krutitsky-04,krutitsky-05,kimura-05,kimura-06,pai-08} it was predicted that the phase transition between the SF and MI states can be of first order at even fillings. 
We model spin-1 bosons in a hypercubic lattice with the following spin-1 Bose-Hubbard Hamiltonian:
\begin{eqnarray}
\hat{H}&=&-t\sum_{\langle j,l\rangle}\sum_{\sigma=0,\pm 1}(\hat{b}^{\dagger}_{j,\sigma} \hat{b}_{l,\sigma} +  {\rm H.c.}) -\mu \sum_j \hat{n}_j\nonumber\\&&+ \frac{U_0}{2}\sum_j \hat{n}_{j}(\hat{n}_{j}-1) + \frac{U_2}{2}\sum_j(\hat{{\bf F}}_j^2-2\hat{n}_j),
\end{eqnarray}
where $\hat{b}^{\dagger}_{j,\sigma}$ creates a boson with spin $\sigma$ at site $j$ of a $D$-dimensional hypercubic lattice, $\hat{n}_{j}=\sum_\sigma \hat{b}^{\dagger}_{j,\sigma}\hat{b}_{j,\sigma}$, $\hat{{\bf F}}_j=\sum_{\sigma,\sigma^\prime}\hat{b}^{\dagger}_{j,\sigma}{\bf F}_{\sigma\sigma^\prime} \hat{b}_{j,\sigma^\prime}$, and ${\bf F}_{\sigma\sigma^\prime}=(F^x_{\sigma\sigma^\prime},F^y_{\sigma\sigma^\prime},F^z_{\sigma\sigma^\prime})$ are the standard $3\times 3$ spin-one matrices~\cite{tsuchiya-04,kimura-05}. This system can be realized using a gas of alkali atoms such as $^{23}$Na, $^{39}$K, and $^{87}$Rb with hyperfine spin $F=1$~\cite{stenger-98,barrett-01,ho-98,ohmi-98,stamperkurn-98,chang-04}. The strength of the interactions $U_0$ and $U_2$ are related to the scattering lengths $a_0$ and $a_2$ for $S=0$ and $S=2$ channels, and it is known that $^{23}$Na has antiferromagnetic interaction $U_2>0$ while $^{87}$Rb is ferromagnetic, $U_2<0$. For antiferromagnetic interaction $U_2>0$, the SF-to-MI transition can be of first order near the tips of the MI lobes at even filling factors~\cite{kimura-05,pai-08}.

\begin{figure}[t]
\includegraphics[scale=0.5]{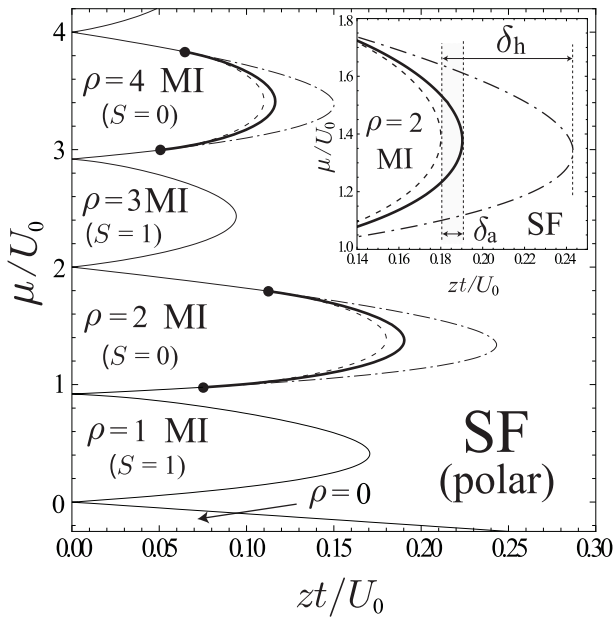}
\caption{\label{spin1}
The ground-state phase diagram of the spin-1 Bose-Hubbard model within the MF approximation for $U_{2}/U_0=0.04$. Second- and first-order phase transitions are indicated by the thin and thick lines, respectively. The dashed (dash-dotted) lines represent the metastability limits of SF (MI) phase, and the dots mark the tricritical points, where the transition changes from first to second order. The inset is an enlarged view of the region around the tip of the $\rho=2$ MI lobe. }
\end{figure}
\begin{figure}[t]
\includegraphics[scale=0.45]{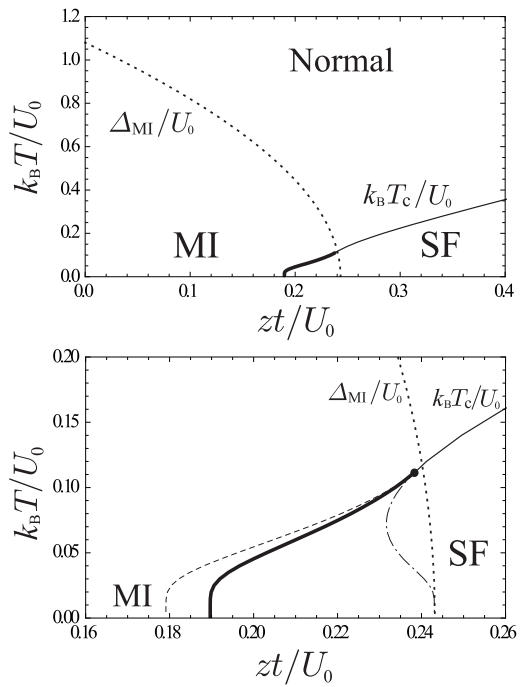}
\caption{\label{fig93}
Finite-temperature phase diagram of the spin-1 Bose-Hubbard model with $U_{2}/U_0=0.04$. The chemical potential $\mu$ is tuned to be the value at the tip of the $\rho=2$ MI lobe; $\mu/U_0=1.338$. Second- and first-order phase transitions are indicated by the thin and thick lines, respectively. We also plot the value of the Mott gap $\Delta_{\rm MI}/U_0$. Lower panel: the enlarged view of the upper panel with adding the metastability limits the SF and MI phases. }
\end{figure}
In Fig.~\ref{spin1}, we show the ground-state phase diagram of the spin-1 Bose-Hubbard model for $U_{2}/U_0=0.04$ obtained by the single-site MF approximation with the decoupling similar to Eq.~(\ref{Decoupling}). This value of $U_{2}/U_0$ corresponds to the experiment with $^{23}$Na atoms~\cite{ho-98,ohmi-98,stamperkurn-98}. For spin-1 bosons, one has to solve the self-consistent equations for three mean fields $\phi_\sigma\equiv \langle \hat{b}_{\sigma}\rangle$ ($\sigma=0,\pm$). It is found that the SF region has a polar solution $\phi_0\neq 0, \phi_\pm= 0$ as the ground state~\cite{tsuchiya-04}. The phase boundaries between SF and MI obtained by the MF decoupling procedure are identical to the ones obtained before by the Gutzwiller approximation~\cite{kimura-05} due to the equivalence of the two single-site approximations. In addition to the second- and first-order boundaries, we also plotted the metastability limits of SF and MI, at which the corresponding local minima of the energy function disappear. When the filling factor $\rho=\langle\hat{n}\rangle$ is even, the MI state consists of $\rho/2$ singlet pairs with $S=0$ while it has spin $S=1$ at odd filling factors~\cite{tsuchiya-04}. Only the case of even filling factors can exhibit first-order SF-to-MI transitions. One has to go beyond the simple single-site approximation in order to discuss more detailed magnetic structures inside the MI phases~\cite{demler-02,imambekov-03,snoek-04}.

The T=0 phase diagram of Fig.~\ref{spin1} with the metastability limits has a very similar structure to the one of the two-component Bose-Hubbard model with repulsive inter-component interaction in Fig.~\ref{fig6}. Furthermore, as shown in the finite-temperature phase diagram (Fig.~\ref{fig93}), the transition between the SF and $\rho=2$ MI phases remains first order until $T/U_1\approx 0.11$ and there is no quantum critical point. Therefore, the system of spin-1 bosons in an optical lattice is another qualified candidate for simulating first-order transition phenomena in cold-atom systems, although it needs to be cooled to sufficiently low temperatures. We can see that the phase diagram in Fig.~\ref{spin1} (similarly to that of Fig.~\ref{fig6}) has the geometry required for the anomalous hysteresis to appear; the Mott lobes at even fillings are surrounded by first-order transition boundary with the SF phase. Therefore, the unidirectional behavior in the hysteresis is expected to be observed also in the spin-1 system upon varying the chemical potential $\mu/U_0$. To obtain the anomalous hysteresis, the optical-lattice depth has to be tuned so that the value of $zt/U_0$ lies within $\delta_{\rm a}$ shown in the inset of Fig.~\ref{spin1}.

\section{\label{5}SUMMARY}
In summary, we have studied first-order transition phenomena of Bose gases trapped in optical lattices, especially focusing on an anomalous hysteresis behavior that was proposed in Ref.~\onlinecite{yamamoto-12a}. First we analyzes the ground-state phase diagram of the hardcore bosons with full dipole-dipole interactions in a triangular lattice~\cite{pollet-10,yamamoto-12a} by means of the CMF+S method~\cite{yamamoto-12b}. 
Our scaled phase diagram is in good agreement with numerical data obtained by QMC calculations~\cite{pollet-10} within the error bars, except that our result predicts the existence of several additional solid phases. We demonstrated that the first-order melting transition can exhibit an anomalous hysteresis behavior upon sweeping the value of the chemical potential, as in the truncated model with only nearest-neighbor interactions~\cite{yamamoto-12a}. In the anomalous hysteresis cycle, once the solid state melts into the SF, it cannot be solidified back again. Unlike in a conventional first-order transition, the hysteresis curve does not form a hysteresis-loop structure. We have also studied binary Bose mixtures with inter-component repulsion loaded into a hypercubic lattice in order to identify the conditions required for the anomalous hysteresis to emerge. This system is completely different from the triangular-lattice dipolar bosons in terms of component degrees of freedom, dimensionality, interparticle interactions, and lattice geometry. Nevertheless, we found a similar unidirectional behavior in the SF-MI transition of the system upon sweeping the value of the chemical potential; the transition from SF to MI cannot occur whereas the transition in the opposite direction is allowed.

A common feature of the two bosonic systems mentioned above is the characteristic geometry of the phase diagram. In both systems, a phase region of lobe shape (the solid phase for the former case and the Mott insulator for the latter case) is surrounded by the first-order boundary. This geometry is indeed the necessary condition for the anomalous hysteresis to emerge. As another example that has such a geometry, we also discussed the system of spin-1 bosons in a hypercubic optical lattice. Moreover, in our recent work~\cite{yamamoto-13}, we found that a spin-dimer system on a triangular lattice can also exhibit the anomalous hysteresis behavior in the first-order magnetic transition induced by controlling the magnetic field instead of the chemical potential. Based on all these examples, we conclude that the anomalous hysteresis is a ubiquitous phenomenon of systems with a phase region of lobe shape that is surrounded by the first-order boundary.

Furthermore, taking the system of binary Bose mixtures with equal intra-component interactions as a simple example, we constructed a sixth-order Ginzburg-Landau theory for the anomalous hysteresis in order to mathematically establish the unconventional first-order transition. Using the perturbative mean-field method, we determined the coefficients $a_2$, $a_4$, and $a_6$ as functions of the microscopic parameters $t$, $U$, $U_{12}$, and $\mu$. We found out that in the anomalous hysteresis, $a_4$ never exceeds $-\sqrt{3a_2a_6}$ and thereby the energy landscape always has minima at finite SF order parameter. Therefore, SF states remain stable over the entire hysteresis region, resulting in the anomalous unidirectional hysteresis.

Assuming binary mixtures of $^{87}$Rb atoms in the two hyperfine states~\cite{anderlini-07, widera-08, trotzky-08, weld-09, gadway-10} confined to a simple-cubic optical lattice, we estimated the required range of the tuning of lattice depth $V_0$ for a clear observation of first-order transition phenomena. It is expected that the estimation shown here should stimulate further experiments in this direction. We stress that the observation of the first-order SF-to-MI transition will significantly expand the applicability of ultracold atomic/molecular systems as a quantum simulator of strongly interacting many-body physics.

\acknowledgments
We thank KAKENHI (25800228) from JSPS (I. D. ) and ARO (W911NF-09-1-0220) (C. SdM.) for support.


\begin{thebibliography}{99}
\bibitem{bloch-12}
I. Bloch, J. Dalibard, and S. Nascimb\`ene,
Nat. Phys. {\bf 8}, 267 (2012).

\bibitem{trotzky-10}
S. Trotzky, L. Pollet, F. Gerbier, U. Schnorrberger, I. Bloch, N. V. Prokof'ev, B. Svistunov, and M. Troyer, 
Nat. Phys. {\bf 6}, 998 (2010). 

\bibitem{anderlini-07}
M. Anderlini, P. J. Lee, B. L. Brown, J. Sebby-Strabley, W. D.Phillips, and J. V. Porto, 
Nature (London) {\bf 448}, 452 (2007).

\bibitem{trotzky-08}
S. Trotzky, P. Cheinet, S. F\"olling, M. Feld, U. Schnorrberger, A. M. Rey, A. Polkovnikov, E. A. Demler, M. D. Lukin, and I. Bloch, 
Science {\bf 319}, 295 (2008).

\bibitem{widera-08}
A. Widera, S. Trotzky, P. Cheinet, S. F\"olling, F. Gerbier, I. Bloch, V. Gritsev, M. D. Lukin, and E. Demler, 
Phys. Rev. Lett. {\bf 100}, 140401 (2008).

\bibitem{weld-09}
D. M. Weld, P. Medley, H. Miyake, D. Hucul, D. E. Pritchard, and W. Ketterle, 
Phys. Rev. Lett. {\bf 103}, 245301 (2009).

\bibitem{gadway-10}
B. Gadway, D. Pertot, R. Reimann, and D.Schneble, 
Phys. Rev. Lett. {\bf 105}, 045303 (2010).

\bibitem{chin-06}
J. K. Chin, D. E. Miller, Y. Liu, C. Stan, W. Setiawan, C. Sanner, K. Xu, and W. Ketterle,
Nature (London) {\bf 443}, 961 (2006).

\bibitem{jordens-08}
R. J\"ordens, N. Strohmaier, K. G\"unter, H. Moritz, and T. Esslinger, Nature (London) {\bf 455}, 204 (2008). 

\bibitem{schneider-08}
U. Schneider, L. Hackerm\"uller, S. Will, T. Best, I. Bloch, T. A. Costi, R. W. Helmes, D. Rasch, and A. Rosch, Science {\bf 322}, 1520 (2008). 

\bibitem{taie-12}
S. Taie, R. Yamazaki, S. Sugawa, and Y. Takahashi,
Nat. Phys. {\bf 8}, 825 (2012)

\bibitem{catani-08}
J. Catani, L. De Sarlo, G. Barontini, F. Minardi, and M. Inguscio, 
Phys. Rev. A {\bf 77}, 011603(R) (2008).

\bibitem{gunter-06} 
K. G\"unter, T. St\"oferle, H. Moritz, M. K\"ohl, and T. Esslinger,
Phys. Rev. Lett. {\bf 96}, 180402 (2006).

\bibitem{ospelkaus-06} 
S. Ospelkaus, C. Ospelkaus, O. Wille, M. Succo, P. Ernst, K. Sengstock, and K. Bongs,
Phys. Rev. Lett. {\bf 96}, 180403 (2006).

\bibitem{sugawa-11} 
S. Sugawa, K. Inaba, S. Taie, R. Yamazaki, M. Yamashita, and Y. Takahashi,
Nat. Phys. {\bf 7}, 642 (2011).

\bibitem{griesmaier-05}
A. Griesmaier, J. Werner, S. Hensler, J. Stuhler, and T. Pfau, 
Phys. Rev. Lett. {\bf 94}, 160401 (2005).

\bibitem{lu-11}
M. Lu, N. Q. Burdick, S. H. Youn, and B. L. Lev, 
Phys. Rev. Lett. {\bf 107}, 190401 (2011).

\bibitem{aikawa-12}
K. Aikawa, A. Frisch, M. Mark, S. Baier, A. Rietzler, R. Grimm, and F. Ferlaino,
Phys. Rev. Lett. {\bf 108}, 210401 (2012).

\bibitem{ni-08}
K.-K. Ni, S. Ospelkaus, M. H. G. de Miranda, A. Pe'er, B. Neyenhuis, J. J. Zirbel, S. Kotochigova, P. S. Julienne, D. Jin, and J. Ye, 
Science {\bf 322}, 231 (2008).

\bibitem{aikawa-10}
K. Aikawa, D. Akamatsu, M. Hayashi, K. Oasa, J. Kobayashi, P. Naidon, T. Kishimoto, M. Ueda, and S. Inouye, 
Phys. Rev. Lett. {\bf 105}, 203001 (2010).

\bibitem{deiglmayr-08}
J. Deiglmayr, A. Grochola, M. Repp, K. Mortlbauer, C. Gluck, J. Lange, O. Dulieu, R. Wester, and M. Weidem\"uller,
Phys. Rev. Lett. {\bf 101}, 133004 (2008). 

\bibitem{sebby-06}
J. Sebby-Strabley, M. Anderlini, P. S. Jessen, and J. V. Porto,
Phys. Rev. A {\bf 73}, 033605 (2006).

\bibitem{folling-07}
S. F\"olling, S. Trotzky, P. Cheinet, M. Feld, R. Saers, A. Widera, and I. Bloch, 
Nature (London) {\bf 448}, 1029 (2007).

\bibitem{yachen-10}
Y.-A. Chen, S. D. Huber, S. Trotzky, I. Bloch, and E. Altman,
Nat. Phys. {\bf 7}, 61 (2011). 

\bibitem{becker-10}
C. Becker, P. Soltan-Panahi, J. Kronj\"ager, S. D\"orscher, K. Bongs, and K. Sengstock, 
New J. Phys. {\bf 12}, 065025 (2010). 

\bibitem{struck-11}
J. Struck, C. \"Olschl\"ager, R. L. Targat, P. Soltan-Panahi, A. Eckardt, M. Lewenstein, P. Windpassinger, and K. Sengstock, 
Science {\bf 333}, 996 (2011). 

\bibitem{soltan-11}
P. Soltan-Panahi, J. Struck, P. Hauke, A. Bick, W. Plenkers, G. Meineke, C. Becker, P. Windpassinger, M. Lewenstein, and K. Sengstock, 
Nat. Phys. {\bf 7}, 434 (2011).

\bibitem{soltan-12}
P. Soltan-Panahi, D. L\"uhmann, J. Struck, P. Windpassinger, and K. Sengstock, 
Nat. Phys. {\bf 8}, 71 (2012).

\bibitem{tarruell-12}
L. Tarruell, D. Greif, T. Uehlinger, G. Jotzu, and T. Esslinger, 
Nature (London) {\bf 483}, 302 (2012). 

\bibitem{jo-12}
G.-B. Jo, J. Guzman, C. K. Thomas, P. Hosur, A. Vishwanath, and D. M. Stamper-Kurn, 
Phys. Rev. Lett. {\bf 108}, 045305 (2012). 

\bibitem{greiner-02}
M. Greiner, O. Mandel, T. Esslinger, T. W. H\"ansch, and I. Bloch, 
Nature (London) {\bf 415}, 39 (2002). 

\bibitem{stoeferle-04}
T. St\"oferle, H. Moritz, C. Schori, M. K\"ohl, and T. Esslinger, 
Phys. Rev. Lett. {\bf 92}, 130403 (2004).

\bibitem{spielman-07}
I. B. Spielman, W. D. Phillips, and J. V. Porto, 
Phys. Rev. Lett. {\bf 98}, 080404 (2007).

\bibitem{altman-03}
E. Altman, W. Hofstetter, E. Demler, and M. D. Lukin
New J. Phys. {\bf 5}, 113 (2003).

\bibitem{kuklov-04}
A. Kuklov, N. Prokof'ev, and B. Svistunov, Phys. Rev. Lett. {\bf 92}, 050402 (2004).

\bibitem{chen-10}
P. Chen and M. F. Yang, Phys. Rev. B {\bf 82}, 180510(R) (2010).

\bibitem{ozaki-12}
T. Ozaki, I. Danshita, and T. Nikuni, 
arXiv:1210.1370v1.

\bibitem{krutitsky-04}
K. V. Krutitsky and R. Graham, 
Phys. Rev. A {\bf 70}, 063610 (2004)

\bibitem{krutitsky-05}
K. V. Krutitsky, M. Timmer, and R. Graham, 
Phys. Rev. A {\bf 71}, 033623 (2005).

\bibitem{kimura-05}
T. Kimura, S. Tsuchiya, and S. Kurihara, 
Phys. Rev. Lett. {\bf 94}, 110403 (2005). 

\bibitem{kimura-06}
T. Kimura, S. Tsuchiya, M. Yamashita, and S. Kurihara, 
J. Phys. Soc. Jpn. {\bf 75}, 074601 (2006).

\bibitem{pai-08}
R. V. Pai, K. Sheshadri, and R. Pandit, 
Phys. Rev. B {\bf 77}, 014503 (2008). 

\bibitem{deforges-11}
L. de Forges de Parny, F. H\'ebert, V. G. Rousseau, R. T. Scalettar, and G. G. Batrouni, 
Phys. Rev. B {\bf 84}, 064529 (2011).

\bibitem{murthy-97}
G. Murthy, D. Arovas, and A. Auerbach, 
Phys. Rev. B {\bf 55}, 3104 (1997).

\bibitem{wessel-05}
S. Wessel and M. Troyer, 
Phys. Rev. Lett. {\bf 95}, 127205 (2005).

\bibitem{boninsegni-05}
M. Boninsegni and N. Prokof'ev, 
Phys. Rev. Lett. {\bf 95}, 237204 (2005).

\bibitem{heidarian-05} 
D. Heidarian and K. Damle, 
Phys. Rev. Lett. {\bf 95}, 127206 (2005).

\bibitem{melko-05}
R. G. Melko, A. Paramekanti, A. A. Burkov, A. Vishwanath, D. N. Sheng, and L. Balents, 
Phys. Rev. Lett. {\bf 95}, 127207 (2005).

\bibitem{sen-08}
A. Sen, P. Dutt, K. Damle, and R. Moessner, 
Phys. Rev. Lett. {\bf 100}, 147204 (2008).

\bibitem{heidarian-10}
D. Heidarian and A. Paramekanti, 
Phys. Rev. Lett. {\bf 104}, 015301 (2010). 

\bibitem{pollet-10}
L. Pollet, J. D. Picon, H. P. B\"uchler, and M. Troyer, 
Phys. Rev. Lett. {\bf 104}, 125302 (2010).

\bibitem{yamamoto-12a}
D. Yamamoto, I. Danshita, and C. A. R. S\'a de Melo, 
Phys. Rev. A {\bf 85}, 021601(R) (2012).

\bibitem{bonnes-11}
L. Bonnes and S. Wessel, 
Phys. Rev. B {\bf 84}, 054510 (2011). 

\bibitem{zhang-11}
X.-F. Zhang, R. Dillenschneider, Y. Yu, and S. Eggert, 
Phys. Rev. B {\bf 84}, 174515 (2011).

\bibitem{danshita-09}
I. Danshita and C. A. R. S\'a de Melo,
Phys. Rev. Lett. {\bf 103}, 225301 (2009).

\bibitem{batrouni-00}
G. G. Batrouni and R. T. Scalettar, 
Phys. Rev. Lett. {\bf 84}, 1599 (2000).

\bibitem{yamamoto-12b} 
D. Yamamoto, A. Masaki, and I. Danshita, 
Phys. Rev. B {\bf 86}, 054516 (2012). 

\bibitem{goral-02}
K. Goral, L. Santos, and M. Lewenstein, Phys. Rev. Lett. {\bf 88}, 170406 (2002). 

\bibitem{fukui-09}
K. Fukui and S. Todo, J. Comput. Phys. {\bf 228}, 2629 (2009). 

\bibitem{sansone-10a}
B. Capogrosso-Sansone, C. Trefzger, M. Lewenstein, P. Zoller, and G. Pupillo,
Phys. Rev. Lett. {\bf 104}, 125301 (2010).

\bibitem{ohgoe-11}
T. Ohgoe, T. Suzuki, and N. Kawashima,
J. Phys. Soc. Jpn. {\bf 80}, 113001 (2011).

\bibitem{jaksch-98}
D. Jaksch, C. Bruder, J. I. Cirac, C. W. Gardiner, and P. Zoller, Phys. Rev. Lett. {\bf 81}, 3108 (1998).

\bibitem{thalhammer-08}
G. Thalhammer, G. Barontini, L. De Sarlo, J. Catani, F. Minardi, and M. Inguscio, 
Phys. Rev. Lett. {\bf 100}, 210402 (2008).

\bibitem{tojo-10}
S. Tojo, Y. Taguchi, Y. Masuyama, T. Hayashi, H. Saito, and T. Hirano, Phys. Rev. A {\bf 82}, 033609 (2010).

\bibitem{mckay-10}
D. McKay and B. DeMarco, New J. Phys. {\bf 12}, 055013 (2010).

\bibitem{chen-03}
G.-H. Chen and Y.-S. Wu,
Phys. Rev. A {\bf 67}, 013606 (2003).

\bibitem{han-04}
J.-R. Han, T. Zhang, Y.-Z. Wang, and W.~M. Liu, 
Phys. Lett. A {\bf 332},  131  (2004).

\bibitem{isacsson-05}
A. Isacsson, M.-C. Cha, K. Sengupta, and S.~M. Girvin,
Phys. Rev. B {\bf 72}, 184507 (2005).

\bibitem{mathey-07}
L. Mathey,
Phys. Rev. B {\bf 75}, 144510 (2007).

\bibitem{arguelles-07}
A. Arg{\"u}elles and L. Santos,
Phys. Rev. A {\bf 75},  053613  (2007).

\bibitem{mathey-09}
L. Mathey, I. Danshita, and C. W. Clark,
Phys. Rev. A {\bf 79}, 011602(R) (2009).

\bibitem{hu-09}
A. Hu, L. Mathey, I. Danshita, E. Tiesinga, C.~J. Williams, and C.~W. Clark, 
Phys. Rev. A {\bf 80},  023619  (2009).

\bibitem{hubener-09}
A. Hubener, M. Snoek, and W. Hofstetter,
Phys. Rev. B {\bf 80}, 245109 (2009).

\bibitem{sansone-10b}
B. Capogrosso-Sansone, {\c{S}}.~G. S{\"o}yler, N.~V. Prokof'ev, and B.~V. Svistunov,
Phys. Rev. A {\bf 81}, 053622 (2010).

\bibitem{iskin-10}
M. Iskin,
Phys. Rev. A {\bf 82}, 033630 (2010).


\bibitem{fisher-89}
M. P. A. Fisher, P. B. Weichman, G. Grinstein, and D. S. Fisher, Phys. Rev. B {\bf 40}, 546 (1989). 

\bibitem{sheshadri-93}
K. Sheshadri, H. R. Krishnamurthy, R. Pandit, and T.V. Ramakrishnan, Europhys. Lett. {\bf 22}, 257 (1993). 

\bibitem{oosten-01}
D. van Oosten, P. van der Straten, and H. T. C. Stoof, Phys. Rev. A {\bf 63}, 053601 (2001). 

\bibitem{buonsante-04}
P. Buonsante and A. Vezzani, Phys. Rev. A {\bf 70}, 033608 (2004).

\bibitem{lu-06}
X. Lu and Y. Yu, Phys. Rev. A {\bf 74}, 063615 (2006). 

\bibitem{sansone-07}
B. Capogrosso-Sansone, N. V. Prokof\'ev, and B. V. Svistunov, Phys. Rev. B, {\bf 75}, 134302 (2007). 

\bibitem{sachdev-97}
S. Sachdev, Phys. Rev. B {\bf 55}, 142 (1997). 

\bibitem{swallowtail}
See, for example, 
E. J. Mueller, 
Phys. Rev. A {\bf 66}, 063603 (2002); 
B. Li, H. Meng, W. Ren, and Z. Zhang, 
Europhys. Lett. {\bf 97}, 57002 (2012); 
B. J. Baelus, F. M. Peeters, and V. A. Schweigert, 
Phys. Rev. B {\bf 63}, 144517 (2001). 

\bibitem{mertes-07}
K. M. Mertes, J. W. Merrill, R. Carretero-Gonz\'alez, D. J. Frantzeskakis, P. G. Kevrekidis, and D. S. Hall, 
Phys. Rev. Lett. {\bf 99}, 190402 (2007). 

\bibitem{bloch-08}
I. Bloch, J. Dalibard, and W. Zwerger, Rev. Mod. Phys. {\bf 80}, 885 (2008). 

\bibitem{tsuchiya-04}
S. Tsuchiya, S. Kurihara, and T. Kimura, 
Phys. Rev. A {\bf 70}, 043628 (2004).

\bibitem{mitra-08}
K. Mitra, C. J. Williams, and C. A. R. S\'a de Melo, 
Phys. Rev. A {\bf 77}, 033607 (2008).

\bibitem{rizzi-05}
M. Rizzi, D. Rossini, G. De Chiara, S. Montangero, and R. Fazio, 
Phys. Rev. Lett. {\bf 95}, 240404 (2005). 

\bibitem{bergkvist-06}
S. Bergkvist, I. P. McCulloch, and A. Rosengren, 
Phys. Rev. A {\bf 74}, 053419 (2006). 

\bibitem{batrouni-09}
G. G. Batrouni, V. G. Rousseau, and R. T. Scalettar, 
Phys. Rev. Lett. {\bf 102}, 140402 (2009). 

\bibitem{stenger-98}
J. Stenger, S. Inouye, D.M. Stamper-Kurn, H.-J. Miesner, A.P. Chikkatur, W. Ketterle Nature {\bf 396}, 345 (1998).

\bibitem{barrett-01}
M. Barrett, J. Sauer, and M.S. Chapman, Phys. Rev. Lett. {\bf 87}, 010404 (2001).

\bibitem{ho-98}
T.-L. Ho, Phys. Rev. Lett. {\bf 81}, 742 (1998).

\bibitem{ohmi-98}
T. Ohmi and K. Machida,J. Phys. Soc. Jpn. {\bf 67}, 1822 (1998)

\bibitem{stamperkurn-98}
D. M. Stamper-Kurn, M. R. Andrews, A. P. Chikkatur, S. Inouye, H.-J. Miesner, J. Stenger, and W. Ketterle, Phys. Rev. Lett. {\bf 80}, 2027 (1998).

\bibitem{chang-04}
M.-S. Chang, C.D. Hamley, M.D. Barrett, J.A. Sauer, K.M. Fortier, W. Zhang, L. You, M.S. Chapman, Phys. Rev. Lett. {\bf 92}, 140403 (2004).

\bibitem{demler-02}
E. Demler and F. Zhou, Phys. Rev. Lett. {\bf 88}, 163001 (2002).

\bibitem{imambekov-03}
A. Imambekov, M. Lukin, and E. Demler, Phys. Rev. A {\bf 68}, 063602 (2003).
 
\bibitem{snoek-04}
M. Snoek and F. Zhou, Phys. Rev. B {\bf 69}, 094410 (2004).

\bibitem{yamamoto-13} 
D. Yamamoto and I. Danshita, Phys. Rev. B {\bf 88}, 014419 (2013). 

\end{thebibliography}
\end{document}